\documentclass[12pt]{article}
\usepackage{graphicx}
\usepackage{amsmath}
\usepackage{slashed}
\usepackage{amsfonts}
\usepackage{amssymb}
\usepackage{mathtools}
\usepackage{makeidx}   
\usepackage{float}
\usepackage{gensymb}
\usepackage{comment}
\usepackage{subcaption}
\usepackage{upgreek}
\usepackage{placeins}
\usepackage{epstopdf}
\usepackage{jheppub} 

\usepackage{xcolor}

\newcommand{\inchsign}{^{\prime\prime}}

\makeatletter
\g@addto@macro\@floatboxreset\centering
\makeatother

\title{Searching for Dark Pions with the MoEDAL-MAPP Detector at the LHC}

\author[a,b]{Shafakat Arifeen,\note{Communicating author.}}
\author[b]{Pierre-Philippe Ouimet,}
\author[a]{James Pinfold,}
\author[a,c]{Michael Staelens}

\affiliation[a]{Department of Physics, University of Alberta, Edmonton, Alberta, T6G 2E1, Canada}
\affiliation[b]{Department of Physics, University of Regina, Regina, Saskatchewan, S4S 0A2, Canada}
\affiliation[c]{Instituto de F\'isica Corpuscular, CSIC--Universitat de Val\`encia, Catedr\'atico Jos\'e Beltr\'an, 2, Paterna, 46980, Val\`encia, Spain}

\abstract{
We investigate the potential for detecting pion-like dark matter within the framework of a Strongly Interacting Massive Particle (SIMP) model using the MoEDAL Apparatus for Penetrating Particles (MAPP) at the LHC. The model, motivated by Chiral Perturbation Theory, features milli-charged dark pions that couple to Standard Model particles via kinetic mixing with a dark photon. Production channels considered include Drell--Yan pair production and photon fusion via the Wess--Zumino--Witten term, the latter providing a distinctive three-body final state signature. Cross-sections are computed using MadGraph and WHIZARD for a range of dark pion masses, decay constants, and dark $Z$ boson masses. Sensitivity projections for MAPP-1 during the High-Luminosity LHC phase are presented under a background-free assumption, showing competitive reach into previously unexplored parameter space for milli-charged dark pseudoscalars. These results highlight the complementary role of MoEDAL-MAPP in probing dark sector models and motivate further detector-level simulations to refine exclusion limits.}


\arxivnumber{2509.02821} 

\begin{document} 
\maketitle
\flushbottom

\section{Introduction}
In the last few decades, strong evidence has been found to support the existence of Dark Matter (DM). Numerous experiments~\cite{Zwicky:1933gu,andernach2017english,1970ApJ...159..379R, Clowe_2006, Spergel_2003} give indirect support to its presence, and recent estimates suggest that it comprises $\sim 25\%$ of the matter/energy density of the Universe. However, the Standard Model (SM) of particle physics has no viable candidate that explains the observed properties of dark matter, and no direct detection of DM has been made so far.

To date, the existing evidence indicates that DM should interact gravitationally. In addition, the data strongly favor two key properties: the amount of Dark Matter (DM) cannot change much; it is generally assumed to be stable, and it is generally considered to be both electrically and effectively neutral compared to the SM\footnote{A subcomponent of DM can be electrically charged, but this charge should be small~\cite{howdarkisdm, xenonDM}.}~\cite{Kribs_2016}. 

However, a weak or feeble nongravitational interaction with SM matter could also help explain the current DM abundances. The cosmic-coincidence problem, which shows that the relic abundances of DM and SM are of the same order of magnitude, could be explained if DM interacted with SM minimally, giving rise to $\textrm{DM} \rightarrow \textrm{SM}$ processes.

Another key issue to consider is that, if DM interacts only gravitationally, as is the case in the $\Lambda$CDM paradigm, DM should have a cuspy central density~\cite{Oh_2008,Oh_2015,cusp1}. However, while simulations using $\Lambda$CDM produce distributions in which the density increases sharply at the center of galaxies, observations of dwarf galaxies show constant central distributions~\cite{deBlok}. This argues for the presence of additional nongravitational interactions.

One of the earliest and best-tested DM theories posits the existence of Weakly Interacting Massive Particles (WIMPs). They are motivated by the assumption that DM interacts with SM via a weak interaction. However, the absence of empirical evidence indicating the existence of WIMPs~\cite{Roszkowski_2018}
has prompted an expansion in the exploration of DM models to encompass alternative candidates that could be generated using existing accelerator technologies. One such class of theories minimally extends the SM to include a dark/hidden sector that is minimally coupled to the SM by interactions called portal interactions~\cite{Holdom:1985ag}. In these theories, DM is assumed to be Feebly Interacting Massive Particles (FIMPs), which have weaker interactions with the SM compared to WIMPs. In certain classes of these models, the $U(1)_Y$ gauge field of the Standard Model is coupled with additional gauge fields that exist in the dark sector (DS). For example, one could have a massless Abelian $U(1)'$ gauge field, with dark photons coupled to the SM hypercharge gauge field $B_\mu$\footnote{The SM hypercharge gauge field is defined by $B_\mu = \cos\theta_W A_\mu - \sin\theta_W Z_\mu$, where $\cos\theta_W$ and $\sin\theta_W$ are the Weinberg angles, and $A_\mu$ and $Z_\mu$ are the photon and $Z$ gauge fields respectively.}. Such a term would be included in the Lagrangian as
\begin{equation}\label{eq:1.a}
    \mathcal{L}_\textrm{mix} = -\frac\kappa 2 A'_{\mu\nu} B^{\mu\nu}
\end{equation}
where $\kappa$ is $< 1$, which governs the amount of mixing between the SM and the Dark Sector (DS). This term is known as a vector portal or kinetic mixing term and is gauge invariant for $U(1)$ gauge fields, as well as being a Lorentz scalar. 

Since a dark $U(1)'$ gauge field, which we will call the dark photon, is introduced, we can also introduce a Dark Quantum Electrodynamics (Dark QED) model. In such a model, dark fermions $\psi_D$, with mass $m_{\psi_D}$, would couple to the dark photon field with a charge of $e'$. Due to the presence of kinetic mixing with $\kappa < 1$, these dark fermions would then become effectively ``milli-charged particles (mCPs)''; i.e., particles with effective electric charges that are a fraction of the charge of the electron $e$. We can see this by performing a field redefinition to diagonalize the mixing term (Eq.~\eqref{eq:1.a}). Using the redefinition $$\partial_\mu \rightarrow \partial_\mu + \dot\imath e' A'_\mu - \dot\imath\kappa e' B_\mu$$
After this redefinition, the Lagrangian becomes
\begin{equation}
    \mathcal{L} = \mathcal{L}_{\textrm{SM}} -\frac14 A'^{\mu\nu}A'_{\mu\nu} + \dot\imath \Bar{\chi}(\slashed{\partial} + \dot\imath e' \slashed{A}' -\dot\imath\kappa e' \slashed{B} + \dot\imath m_\chi)\chi
\end{equation}
where $\chi$ is the dark fermion, and $\slashed{A}' \equiv \gamma^\mu A'_\mu$.
\\Note that the particle $\psi$ now couples to the SM hypercharge gauge field with a charge $\kappa e'$. For vector portal interactions, the charged dark matter fields that couple to the dark photon acquire an effective electric charge due to kinetic mixing. This electric charge can be a small fraction of that of the electron, making these milli-charged particles.

For WIMPs, the scenario most well studied has the DM relic abundance set by annihilation of $2\rightarrow 2$ into SM particles. However, this scenario is not the only way to generate the observed DM relic abundance. We can consider QCD-like models where the DM is strongly self-interacting. In such Strongly Interacting Massive Particle (SIMP) models, the observed DM relic abundance can use a $3\rightarrow 2$ annihilation process that reduces the number of DM particles to obtain the observed abundance \cite{Kribs_2016}. 

As this category of models is QCD-like, it includes a confinement scale that suppresses the self-interactions of the dark mesons and baryons found in the theory. SIMP models can also provide DM candidates that are stable and neutral for SM interactions. Finally, they can also generate a large number of additional observables, leading to a rich stable of hidden sector particles that could potentially be produced and observed.

In this paper, we will consider a SIMP model that is motivated by Chiral Perturbation Theory (ChPT), an effective field theory of QCD. Our model focuses on meson-like DM and couples this DM model to the SM via the kinetic mixing term introduced above. As a result, the charged dark mesons become effectively milli-charged when interacting with SM matter.

For this model we assume that there are at least three light dark quark flavours, which means that our dark QCD has an approximate chiral $SU(3)_{L}\times SU(3)_{R}$ chiral symmetry that is explicitly broken by the dark quark mass terms. As long as the mass of the light dark quarks is small compared to dark QCD confinement scale this can be used as a valid symmetry of the theory. This symmetry is then spontaneously broken to the $SU(3)_{\rm Vector}$ subgroup, and the dark pions are the (approximate) Goldstone bosons of this broken symmetry. Their mass then entirely driven by the explicit breaking of the initial chiral symmetry by the dark quark mass terms. 

 The milli-charged dark pions in this model are stable, and therefore cannot constitute the bulk of dark matter, as strong CMB constraints limit milli-charged dark matter to a maximal subcomponent of ~0.4\% for non-negligible charges \cite{Kovetz_2018,Boddy_2018,de_Putter_2019}. A more general upper limit was obtained by PandaX \cite{Kovetz_2018} under the assumption that 100\% of dark matter is milli-charged, but under that assumption the effective charge is limited to a value below $2.6 \times 10^{-11} \: e$. This is far below the detection threshold that can be achieved by any proposed direct-detection experiment.
 
Therefore, stable dark matter in this model must be made up mainly of neutral stable particles. In many similar models considered in the literature, the lightest neutral dark pion is taken to be stable, forming the bulk of dark matter. In this work, we consider photon fusion as a production process for our dark pions. This means that we must also allow the dark pion to decay into dark photons and SM photons. For standard model pions the decay $\pi^{0} \rightarrow \gamma \gamma$ is the dominant decay mode and happens through the chiral anomaly. If we consider similar behavior for the dark pions, they cannot form the bulk of dark matter. The simplest alternative candidate would be the dark neutrons under the assumption that they are stable. Possible stable dark baryons have been studied in the literature \cite{Bernreuther_2020,Braat_2023, Kamada:2021cow} and we note that no constraints require the mass of the dark quarks or their dark charges to be the same as those of the SM quarks. It is therefore entirely possible for the dark quark masses and charges to be arranged in such a way that the dark neutron, not the dark proton, is stable. This could form a viable dark matter candidate if the dark pions are unstable. In Ref.\cite{Kamada:2021cow} for example, the authors consider a dark neutron model with a range of mixing parameters that are similar to those used in this work.

It is also possible to allow the dark pion to decay to dark photons while still contributing the bulk of dark matter in the early universe as long as they are sufficiently long lived \cite{Berlin_2018}. To do so, their decay to two dark photons must be heavily suppressed. This is done by a judicious choice of the dark quark charges as will be discussed in section 3. This also would not allow for experimentally observable rates of photon fusion, but would not change the Drell--Yan results we obtain. 

The main focus of our current study focuses on dark pions; similar models have been well studied in the literature \cite{Cheng_2022,Kribs_2019,Braat_2023}. However, these models involve either purely massless or purely massive dark gauge fields. Massless dark gauge fields are also frequently used to benchmark experimental searches \cite{Fabbrichesi_2021,Lagouri:2022ier}. The model we study in this work proposes both a massive dark gauge field (a dark Z) and a massless dark photon. Such a combination could easily arise from a dark sector with an $SU(3)\times SU(2)\times U(1)$ symmetry group, where the $SU(2)$ symmetry is spontaneously broken along the same general pattern as in the standard model. It is important to note that we can adjust how much the massive and massless dark gauge fields contribute by adjusting the value of the dark Weinberg angle. This allows us the flexibility consider different amounts of massive and massless gauge fields in future work. For this work we set the dark Weinberg angle to be the same as the SM Weinberg angle. We are not aware of any systematic studies of the cosmological consequences of identical models. Nonetheless, we can look at existing limits for massive dark gauge field to get an idea of the possible parameter space that is cosmologically relevant. 

For example in \cite{Berlin_2018}, they present a model that has long lived dark pions forming the bulk of dark matter in the early universe. It is also important to note that they explicitly consider the influence of dark vector mesons on the behaviour of the model. For this model, they find that dark pions in mass ranges from 10 MeV to 10 GeV with kinetic mixing parameters between $10^{-7}$ and $10^{-1}$ match the observed dark matter energy density. With heavier dark pions they observe larger kinetic mixing parameters being needed, while lighter dark pions require smaller mixing parameters.  

We will investigate the detectability of this model's dark pion by MoEDAL's (Monopole and Exotics Detector at the LHC)  MAPP (MoEDAL Apparatus for Penetrating Particles) subdetector. MoEDAL is the seventh experiment at the Large Hadron Collider (LHC)~\cite{Design, FairbairnPinfold}, while MAPP-1 has been installed adjacent to the MoEDAL detector to take data during the High-Luminosity (HL) phases of the LHC.  

The physics program of the MoEDAL experiment is broad, but concentrates on the search for avatars of new physics. MoEDAL proper is dedicated to the search for highly ionizing avatars of new physics, expanding the physics reach of the LHC in a complementary way. Magnetic monopoles are the main example of such highly ionizing particles, but MoEDAL can also search for any massive, (pseudo-)stable, slow-moving particles with single or multiple electric charges, as found in many BSM scenarios. This aspect of MoEDAL's physics program is described in detail elsewhere~\cite{Acharya_2014}. To date, MoEDAL has taken data in proton--proton collisions with $8$~TeV and $13$~TeV collision energies, as well as in heavy-ion collisions. MoEDAL's most recent limits on monopole production at the LHC are given in Ref.~\cite{acharya-PhysRevLett.123.021802}. 

 MoEDAL's physics program also extends to the search for minimally ionizing particles and long-lived neutral particles. This aspect of the physics program utilizes the MAPP detector, which is specifically designed to detect such particles. The phase-1 MAPP detector (MAPP-1) is currently being installed in the UA83 gallery adjacent to the MoEDAL/LHCb region at Interaction Point~8 (IP8). 

The remainder of this paper is organized as follows. Section~\ref{Section:MAPP} provides a detailed overview of the MAPP-1 detector. Section~\ref{Section:Model} introduces and further motivates the model used in this study. Section~\ref{Section:Results} describes our simulations and presents the results for the processes considered. Finally, Section~\ref{Section:Conclusions} offers concluding remarks.

\section{The MAPP-1 Detector}
\label{Section:MAPP}

\begin{figure}[htbp]
    \centering
    \includegraphics[width=0.75\linewidth]{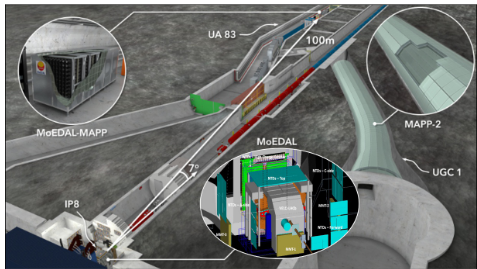}
    \caption{The layout of the MoEDAL-MAPP facility near IP8, including the MoEDAL detector (IP8), MAPP-1 (UA83), and the planned MAPP-2 (UGC1), reproduced with permission from Ref. \cite{Kalliokoski:2023cgw}.}
    \label{MAPP}
\end{figure}

An earlier concept presented alongside the original 1999 MoEDAL proposal envisioned an active detector system targeting exotic, highly penetrating, long-lived particles at the LHC~\cite{PINFOLD199952}. This idea has since materialized in the form of MAPP-1---a fully active scintillation detector with a sensitive volume of approximately $3$~m$^{3}$ installed in the UA83 tunnel near Interaction Point 8 (IP8). The detector is aligned to point back at the IP, positioned roughly $97.8$~m from the IP and angled at about $7.3^{\circ}$ relative to the beamline. The UA83 site offers approximately $110$~m of rock overburden, significantly reducing the background from cosmic radiation, along with around  $50$~m of intervening shielding material between the detector and the IP. A schematic of the MoEDAL-MAPP facility is shown in Figure~\ref{MAPP}.

\begin{figure}[htbp]
    \centering
    \includegraphics[width=0.6\linewidth]{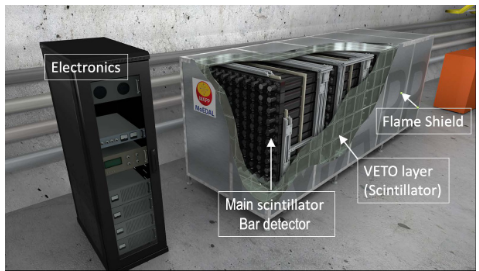}
    \caption{A schematic of MAPP-1 with its primary components highlighted, reproduced with permission from Ref. \cite{Kalliokoski:2023cgw}.}
    \label{MAPP1}
\end{figure}

The MAPP-1 detector consists of four collinear sections, each containing $100$ scintillator bar units measuring $10$~cm $\times$ $10$~cm $\times$ $75$~cm and individually connected to a single low-noise, high-gain $3.1\inchsign$ photomultiplier tube (PMT). This four-layer setup allows coincidence detection that suppresses false signals from PMT dark noise to negligible levels. Each unit comprises two blue-emitting plastic scintillator bars ($5$~cm $\times$ $10$~cm $\times$ $75$~cm; SP32, NUVIATech Instruments, CZ~\cite{KAPLON2023168186}), which were doped for improved light output. Scintillator and PMT testing was performed at the University of Alberta, where detector material preparation and assembly took place in a Grade-C cleanroom. Bars were wrapped and taped to ensure light-tightness: they were first polished, then covered with two layers of Tyvek\textsuperscript{\textregistered}, one layer of black paper, and two layers of black electrical tape. A silicone light guide, made in-house using SYLGARD™~184, was attached to one end of the assembled scintillator bar unit, with a PMT fixed to it. An LED, secured with optical epoxy, was embedded in the opposite end for calibration. The detector frame was built from T-slotted aluminum and high-density polyethylene, with slots designed to hold the scintillator bar units securely in place. Finally, MAPP-1 is enclosed by a hermetic veto system composed of $25$~cm $\times$ $25$~cm $\times$ $1$~cm plastic scintillator subplanes, each containing two embedded fast wavelength-shifting fibers read out by silicon photomultipliers (SiPMs); for safety, the detector is fully encased in an aluminum flame shield. The readout electronics are located adjacent to the MAPP-1 detector in the UA83 gallery. Figure~\ref{MAPP1} presents a schematic of the MAPP-1 detector.

The front-end electronics are designed to accommodate a broad range of signal amplitudes with high precision. The analog-to-digital converter (ADC) achieves an effective dynamic range of approximately $13$~bits ($80$~dB), enabling direct digitization of signals across a wide amplitude spectrum. An integrated programmable-gain amplifier (PGA) further extends the dynamic range by an additional $31.5$~dB. Calibration will be conducted using muons at low gain settings, while operation at higher gain will result in ADC saturation for muons. In such cases, muon signals will be indirectly quantified using the time-over-threshold method. The typical signal rise time lies between $10$ and $20$~ns. The muon calibration will be carried to the mCP level using the LED pulser system, that can simulate the light levels expected by varying the LED pulse and amplitude.

Initial low-level triggering is handled by field-programmable gate arrays (FPGAs), which include per-channel buffers capable of storing $100$~$\upmu$s of raw, unsuppressed data, equivalent to approximately $400$ bunch crossings. This buffer depth ensures sufficient time for trigger decisions and subsequent data transfer to memory. Communication between detector modules is supported via a dedicated backplane, enabling more advanced trigger coordination across channels. While the system emphasizes a flexible software-based triggering approach, several predefined trigger categories are also implemented. One such trigger, essential for mCP detection, requires a quadruple coincidence of photoelectron signals---one in each of the four collinear detector segments---indicating the passage of an mCP with sufficient energy deposition.

All data are initially stored on a local server housed in the UA83 gallery and connected via Ethernet, and subsequently transferred to multiple off-site machines with redundant storage for backup and analysis. Events flagged by initial triggers are sent to shared memory ($8$~GB), where a CPU performs further analysis and selects events for final readout via Ethernet.

\newpage
\section{The Milli-Charged Scalar Model}\label{Section:Model}

Ref.~\cite{Hochberg_2014} introduces a SIMP model that uses a $3\rightarrow 2$ interaction to obtain the observed DM relic abundance. A term involving such an interaction occurs naturally in the QCD sector of the Standard Model and is known as the Wess--Zumino--Witten (WZW) term~\cite{WESS197195,WITTEN1983422}. However, a similar term would be found in any model whose quark structure is identical to that of the SM~\cite{Donoghue_Golowich_Holstein_2023}.

 The alternative mechanism used in the SIMP model considered produces the observed thermal relic DM abundance even if the effective coupling constant of the $2\rightarrow2$ annihilation cross-section is $a_{\textrm{ann}} \simeq 0$. Because of this, the model allows for the DM mass scale to be at or below the GeV scale. Indeed, the mass scale for this mechanism is given by
\begin{equation}
    m_{\textrm{DM}}\sim a_{\textrm{eff}}(T^2_{\textrm{eq}}M_{\textrm{Pl}})^{1/3}\sim 100 \textrm{ MeV}
\end{equation}
Here, $a_{\textrm{eff}}$ is the effective strength of the $3\to2$ self-interaction.

This mechanism, the Strongly Interacting Massive Particle mechanism, therefore predicts light-dark matter with masses in the MeV to GeV range with strong self-interactions. 

Further, we are interested in studying dark pions that are milli-charged, as these are of interest for our collider mCP searches. While we motivate our model by using Chiral Perturbation Theory (ChiPT), we note that the lowest-order terms of ChiPT mostly represent a generic model of pseudo-scalars with the usual kinetic term as well as standard 4-meson self-interaction terms. We therefore motivate our theory using this model, but the Pion fields are taken as a part of the Dark Sector, the confinement scale is potentially much higher than that of QCD, and the underlying strongly interacting theory can be substantially different in other ways as well. 

Ref.~\cite{Hochberg_2015} gives us a model-dependent limit on the abundance of Pion-like Dark Matter:
\begin{equation}\label{eq:2.1.0.5}
    \frac{m_{\pi_D}}{F} \lesssim 2\pi
\end{equation}
where $F$ is the dark pion-decay constant and $m_{\pi_D}$ is the mass of the dark pion. Note that there are substantial differences between the model used in~\cite{Hochberg_2015} and the one considered here. This model does not use a vector portal to couple their hidden sector to the SM. As such, we expect that there should be differences. A similar limit is, however, also discussed in \cite{Berlin_2018} for a similar model that kinetically mixes a single massive dark gauge field with the SM. This expression provides a rough upper limit on the dark pion mass. Since $F$ is a free parameter of the model\footnote{$F$ depends on the confinement scale of the theory since the only restriction of a Pion-like DM theory is $m_q << \Lambda$, where $m_q$ is the mass of the lightest dark quarks and $\Lambda$ is the confinement scale, we can set an arbitrarily high confinement scale and treat $F$ as a free parameter.}, we can set a high value of $F$ if we wish to consider higher value of $m_{\pi_D}$, as long as the ratio satisfies Eq.~\eqref{eq:2.1.0.5}.

 Ref.~\cite{Hochberg_2015} has also stated a lower bound on the mass of the dark pions based on bullet-cluster constraints as found in~\cite{Clowe_2004,Markevitch_2004,Randall_2008} as well as constraints from halo shapes found in~\cite{Rocha_2013,Peter_2013}:
\begin{equation}
    \frac{\sigma}{m_{\pi_D}} \lesssim 1 \textrm{ cm}^2/\textrm{g} \simeq 1.78\times 10^{12} \textrm{ pb}/\textrm{GeV}
\end{equation}
This lower bound is always maintained in our simulations, as our cross-sections are at most of the order $10^4$~pb.

To build up the Pion-like Dark Matter model, we have to construct the Lagrangian. We define a field $U(x)$ in terms of the $SU(3)$ matrix:
\begin{equation}\label{eq:2.1.1}
    U(x) = \exp\Big(\dot\imath\frac{\phi(x)}{F_0}\Big)
\end{equation}
where $\phi$ is defined as
\begin{equation}\label{eq:2.1.1.5}
    \phi = \sum_{a=1}^8 \phi_a\lambda_a = \begin{pmatrix}\pi^0+\frac{1}{\sqrt{3}}\eta & \sqrt{2}\pi^+ & \sqrt{2}K^+ \\ \sqrt{2}\pi^- & -\pi^0 + \frac{1}{\sqrt{3}}\eta & \sqrt{2}K^0 \\ \sqrt{2}K^- & \sqrt{2}\Bar{K}^0 & -\frac{2}{\sqrt{3}}\eta\end{pmatrix}
\end{equation}
where $\lambda_a$ are the Gell-Mann matrices and $\phi_a = \frac12\mathrm{Tr}[ \lambda_a,\phi]$.
\\
Since we're dealing with pseudoscalars, the most general, chirally invariant Lagrangian one can have is~\cite{scherer2002introduction}
\begin{equation}
    \mathcal{L}_{\textrm{eff}} = \frac{F_0^2}{4}\mathrm{Tr}(\partial_\mu U \partial^\mu U^\dagger)
\end{equation}
where $F_0$ is the pion-decay constant in the chiral limit. Introducing symmetries and external fields gives a Lagrangian of the form
\begin{equation}\label{eq:2.1.3}
    \mathcal{L}_2 = \frac{F_0^2}{4}\mathrm{Tr}[D_\mu U (D^\mu U)^\dagger] + \frac{F_0^2}{4}\mathrm{Tr}(\chi U^\dagger + U \chi^\dagger)
\end{equation}
with the covariant derivative defined in terms of the photon and $Z$ as
\begin{equation}\label{eq:2.1.6}
    \begin{aligned}
        D_\mu U &= \partial_\mu U + \dot\imath l_\mu U - \dot\imath U r_\mu \\
        &= \dot\imath g Q U (A_\mu - \tan\theta_W Z_\mu) + \frac{\dot\imath g\tau_3 U}{\sin\theta_W \cos\theta_W} Z_\mu \\
        & - \dot\imath g U Q(A_\mu - \tan\theta_W Z_\mu)
    \end{aligned}
\end{equation}

Upon constructing the Lagrangian, one might notice that the fields in Eq.~\eqref{eq:2.1.3} are even intrinsic parity because they contain an even number of Goldstone bosons only. We can see this if we assume that there are no external fields except for $\chi$ (``pure QCD case''), the two Lagrangians in the above equations are invariant under $\phi(x)\rightarrow-\phi(x)$. However, Eq.~\eqref{eq:2.1.6} fails to describe processes such as $K^+ K^- \rightarrow \pi^+ \pi^- \pi^0$, as well as the anomalous decay of $\pi^0 \rightarrow \gamma\gamma$ when coupled to electromagnetic fields.

 To form a term that explains the above processes, we look at the effective Wess--Zumino--Witten (WZW) action~\cite{WESS197195, WITTEN1983422}.
This is motivated by the anomalous Ward identities~\cite{adler-1969, adler-berdeen, osti_4045871, bardeen-1969, Bell:1969ts},
 which gives rise to a particular form of the variation of the generating functional~\cite{GASSER1984142, WESS197195}.
 Wess and Zumino have derived this consistency satisfied by the anomalous Ward identities, and they constructed a functional (the action). This interaction Lagrangian cannot be obtained from the chirally invariant Lagrangian. It has been modified by Witten to include the lowest-order equation of motion (EOM), the simplest term possible, which breaks the symmetry of having only an even number of Goldstone bosons~\cite{WITTEN1983422}. 

 As mentioned before, inclusion of the WZW term is important, as it allows $3\rightarrow 2$ annihilation of DM particles in the early universe. However, care must be taken as late time annihilation of dark pions into SM photons can cause observable distortions of the cosmic microwave background. In this work, we wish to provide a benchmark for the detectability of dark pion production via photon fusion at MoEDAL-MAPP. We therefore are interest in the range of parameters for which this can be done. These values, however, are far above reasonable cosmological thresholds. For this work, we further only consider contributions coming from a massless dark photon that kinetically mixes with the SM photon. Effectively we embed $U(1)_{D}$ in the unbroken subgroup $SU(3)_{\rm vector}$ of the global chiral symmetry.  
 
After field redefinitions, to leading order in the pion fields, we get the following Lagrangian:
\begin{equation}\label{eq:2.2.2.5}
    \mathcal{L}_{\textrm{WZW}}^0 = \frac{2 N_c}{15\pi^2 F^5}\epsilon^{\mu\nu\rho\sigma}\mathrm{Tr}(\phi\partial_\mu\phi\partial_\nu\phi\partial_\rho\phi\partial_\sigma\phi)
\end{equation}
In the presence of external fields, and considering only coupling to the photon, $r_\mu = l_\mu = -e A_\mu Q$, Eq.~\eqref{eq:2.2.2.5} is simplified due to the terms involving three and four electromagnetic fields vanishing upon contracting with the Levi--Civita tensor $\epsilon^{\mu\nu\rho\sigma}$, and we obtain
\begin{equation}
\begin{aligned}\label{eq:2.2.4}
    n \mathcal{L}_{\textrm{WZW}}^{\textrm{ext}} &= -en A_\mu J^\mu + \dot\imath\frac{ne^2}{48\pi^2}\epsilon^{\mu\nu\rho\sigma}\partial_\nu A_\rho A_\sigma \\ &\times\mathrm{Tr}[2Q^2(U\partial_\mu U^\dagger - U^\dagger\partial_\mu U) - QU^\dagger Q \partial_\mu U + QUQ\partial_\mu U^\dagger]
\end{aligned}
\end{equation}
where 
\begin{equation}
    J^\mu = \frac{\epsilon^{\mu\nu\rho\sigma}}{48\pi^2}\mathrm{Tr}(Q\partial_\nu U U^\dagger\partial_\rho U U^\dagger\partial_\sigma U U^\dagger + Q U^\dagger\partial_\nu U U^\dagger \partial_\rho U U^\dagger\partial_\sigma U)
\end{equation}
\\

Similar models to the one discussed in this work have been used in the literature, but using only massive dark photons. To avoid depleting the dark pion abundance in the early universe due to their decay into less massive dark photons these models suppress this decay rate by choosing 
\begin{eqnarray*}
    Q={\rm diag}(+1,-1, -1)
\end{eqnarray*}
as their dark quark charge matrix. In this work, we follow the standard model and use

\begin{eqnarray*}
    Q={\rm diag}(+\frac{2}{3},-\frac{1}{3}, -\frac{1}{3}).
\end{eqnarray*}
This choice of $Q$ is also made in, for example, Ref. \cite{Kamada:2021cow}. The choice of $Q$ is important because the chiral anomaly that drives the two photon decay is proportional to $\textrm{Tr}(T_{\pi}Q^{2})$, where $T_{\pi}$ is the $SU(3)$ generator corresponding to the neutral pion. The first choice heavily suppresses this decay while the one we use does not. For our Drell--Yan results, using either choice does not significantly change our results.

To investigate Drell--Yan production of milli-charged dark pions, we introduce new massless $U(1)'$ and massive $SU(2)_{W}'$ fields, $A'_\mu$ and $Z'_\mu$, and include the following term in the gauge Lagrangian:
\begin{equation}\label{eq:3.1.0}
\mathcal{L}_{\textrm{mix}} = -\frac{\kappa}{2}B'_{\mu\nu}B^{\mu\nu}
\end{equation}
As per the Standard Model,
\begin{equation}
\begin{aligned}
    Z_\mu &= \cos\theta_W W_{3\mu} - \sin\theta_W B_\mu \\
    A_\mu &= \sin\theta_W W_{3\mu} + \cos\theta_W B_\mu
\end{aligned}
\end{equation}
We follow the same definitions for the dark sector:
\begin{equation}
\begin{aligned}
    Z'_\mu &= \cos\theta_W' W'_{3\mu} - \sin\theta_W' B'_\mu \\
    A'_\mu &= \sin\theta_W' W'_{3\mu} + \cos\theta_W' B'_\mu
\end{aligned}
\end{equation}
Here, $\cos\theta_W'$ and $\sin\theta_W'$ include a dark sector Weinberg angle, which depend on the dark sector gauge couplings $g_1', g_2'$, such that $\sin\theta_W' = g_1'/\sqrt{g_1'^2 + g_2'^2} \equiv s'$, and $\cos\theta_W' = g_2'/\sqrt{g_1'^2 + g_2'^2} \equiv c'$. Similarly, $c$ and $s$ use the same definition but for the SM Weinberg angle. Here, the dark vector field has a mass term of the form $\frac12 M_2^2 Z'_\mu Z'^\mu$, where $M_2 = M_{W'\pm} / \cos\theta_W'$.

 We perform field re-definitions to diagonalize the mixing term in Eq.~\eqref{eq:3.1.0}. This redefinition is standard~\cite{Izaguirre_2015} as it simplifies the calculations and allows us to treat these processes in \textsc{MadGraph}, our main Monte Carlo event generator. Although this would mix the dark photon and $Z'$ in the WZW term, we omit the $Z'$ in the WZW term (Eq.~\eqref{eq:2.2.4}) of our \textsc{MadGraph} model. The associated processes are suppressed compared to the photon interactions and are not of interest to us at this time. After performing field re-definitions, we get the following term for the dark covariant derivative:

\begin{equation}\label{eq:3.1.1}
\begin{aligned}
D_\mu \phi & = \partial_\mu \phi + \dot\imath g_D Q\phi \Bigg[A'_\mu - \kappa cc' A_\mu + \Big(\kappa s c' + \frac12\kappa s s't' \mathbf{M} +\frac12 \kappa s s' t' - (\kappa c s')^2 t' \Big) Z_\mu \\ 
&+ \Big( \frac12 (\kappa s)^2 s' c' \mathbf{M} - \frac{(\kappa s)^2 c'}{2} - (\kappa c)^2 c' s' - t' + \frac18 (\kappa s s')^2 t' \mathbf{M} + \frac{(\kappa s s')^2 t'}{4} \mathbf{M} - \frac{3 (\kappa s s')^2 t'}{8}\Big) Z'_\mu\Bigg] \\
& g_D \lambda_3\Bigg[\Big(\frac{3 (\kappa s)^2 s'}{8 c'} - \frac{\kappa s}{2 c'}\mathbf{M} - \frac{\kappa s}{ 2 c'} - \frac{(\kappa c)^2 s'}{2 c'}\Big)Z_\mu \Bigg]  \\ 
& +\Big(\frac{1}{s' c'} - \frac{(\kappa s)^2 s'}{8 c}\mathbf{M}^2 -\frac{(\kappa s)^2 s' }{4 c'}\mathbf{M}\Big) Z'_\mu\Bigg] \\
& -\dot\imath g_D \phi Q \Bigg[A'_\mu - \kappa_{cc'} A_\mu + \Big(\kappa s c' + \frac12\kappa s s't' \mathbf{M} +\frac12 \kappa s s' t' - (\kappa c s' t')^2 \Big) Z_\mu \\ 
&+ \Big( \frac12 (\kappa s)^2 s' c' \mathbf{M} - \frac{(\kappa s)^2 c'}{2} - (\kappa c)^2 c' s' - t' + \frac18 (\kappa s s')^2 t' \mathbf{M} + \frac{(\kappa s s')^2 t'}{4} \mathbf{M} - \frac{3 (\kappa s s')^2 t'}{8}\Big) Z'_\mu\Bigg] \\ &\equiv \mathcal{D}_\mu \phi 
\end{aligned}    
\end{equation}
where
\begin{equation}\label{eq:3.1.2}
    \mathbf{M} = \frac{M_1^2 + M_2^2}{M_1^2 - M_2^2}   
\end{equation}

\section{Results} \label{Section:Results}

\begin{figure}[htbp]
    \centering
    \includegraphics[scale = 0.20]{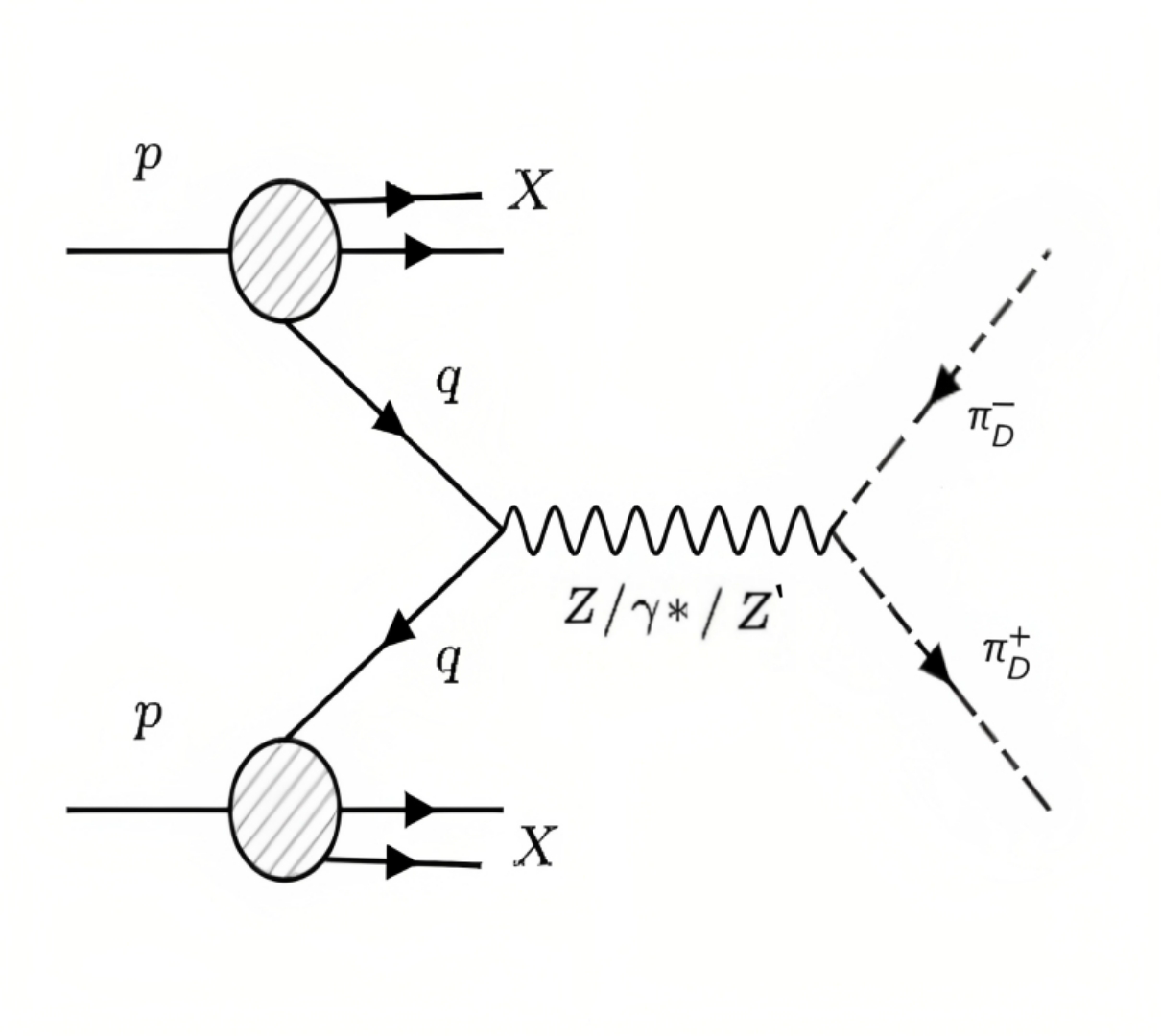}
    \caption{Tree-level Feynman diagram for Drell--Yan pair-produced charged dark pions.}
    \label{DY}
\end{figure}

\begin{figure}[htbp]
    \centering
    \includegraphics[scale = 0.25]{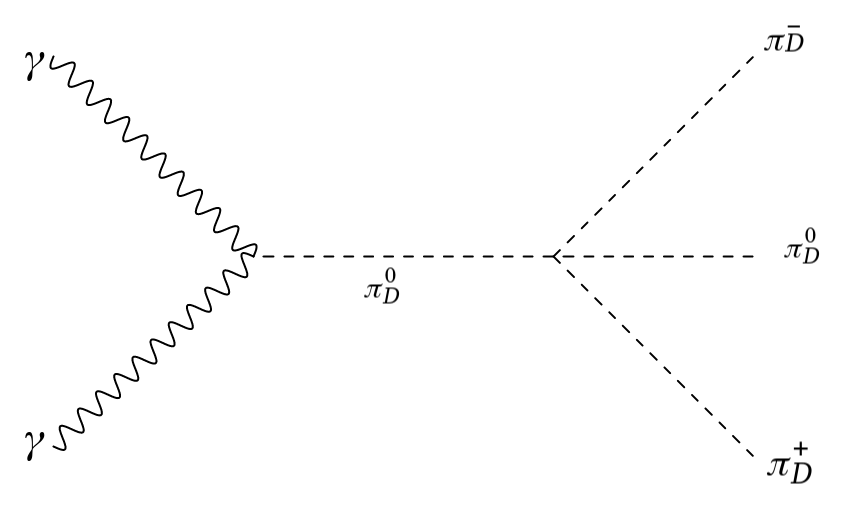}
    \caption{Feynman diagram for \( \gamma \gamma \to \pi_D^0 \pi_D^+ \pi_D^- \) process.}\label{PF}
\end{figure}

\begin{figure}[htbp]
    \centering
    \includegraphics[scale = 0.8]{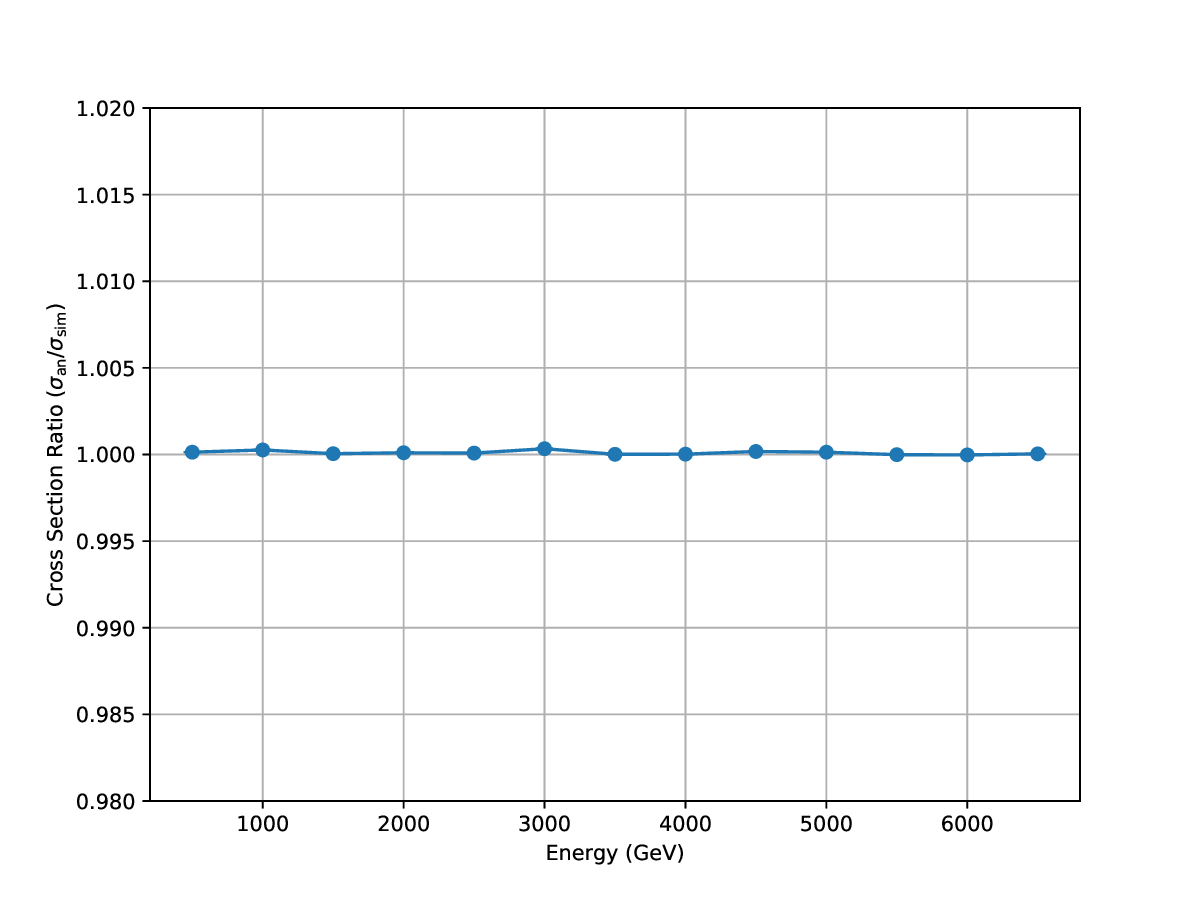}
    \caption{Ratio of $\sigma_{s} / \sigma_{a}$ vs. beam energy $E$ (GeV) for $\pi_D^+ \pi_D^- \rightarrow \pi_D^0 \pi_D^0$. We can see that the ratio is $1$ within error, thus validating the four meson interaction term for our model. It is also useful to note that the points have small error bars that are obscured by the point markers on the plot, corresponding to the errors in \textsc{MadGraph} simulations.}\label{ratio1}
\end{figure}

We study two processes: Drell--Yan pair production of charged dark pions, in Figure~\ref{DY}, and photon fusion to three dark pions, one neutral and two charged, shown in Figure~\ref{PF}. The Drell--Yan process depends on the kinetic term, which includes the covariant derivative, and so we have $A_\mu$, $Z_\mu$, and $Z'_\mu$ as propagators, the latter of which is derived from field re-definitons of the kinetic mixing term in Eq.~\eqref{eq:3.1.0}. The photon-fusion process comes from the WZW term in Eq.~\eqref{eq:2.2.4}, as we see a consequence of the WZW term analogous to the Standard Model: The anomalous decay of a pion to two photons. We used \textsc{MadGraph}~\cite{Alwall_2011} to simulate the Drell--Yan process, and \textsc{WHIZARD}~\cite{Kilian_2011} to simulate the photon-fusion process. This also allows us to cross-check results between two different matrix element calculators.
Before simulating the processes, we must ensure that the kinetic term in Eq.~\eqref{eq:2.1.3} is implemented correctly. We calculate the cross-section of the process $\pi_D^+ \pi_D^- \rightarrow \pi_D^0 \pi_D^0$ analytically, which gives 
\begin{equation}
     \sigma_a = \frac{E^2}{8\pi F^4}
\end{equation}
We then estimate the cross-section ($\sigma_s$) with \textsc{MadGraph} and obtain a ratio plot of the two cross-sections, shown in Figure~\ref{ratio1}.
We perform similar work for $K^+_D K^-_D \rightarrow K^+_D K^-_D$, the analytical cross-section is
\begin{equation}
    \sigma_a = \frac{E^2}{12 \pi F^4}
\end{equation}
Then, the ratio plot is
\begin{figure}[htbp]
    \centering
    \includegraphics[scale = 0.8]{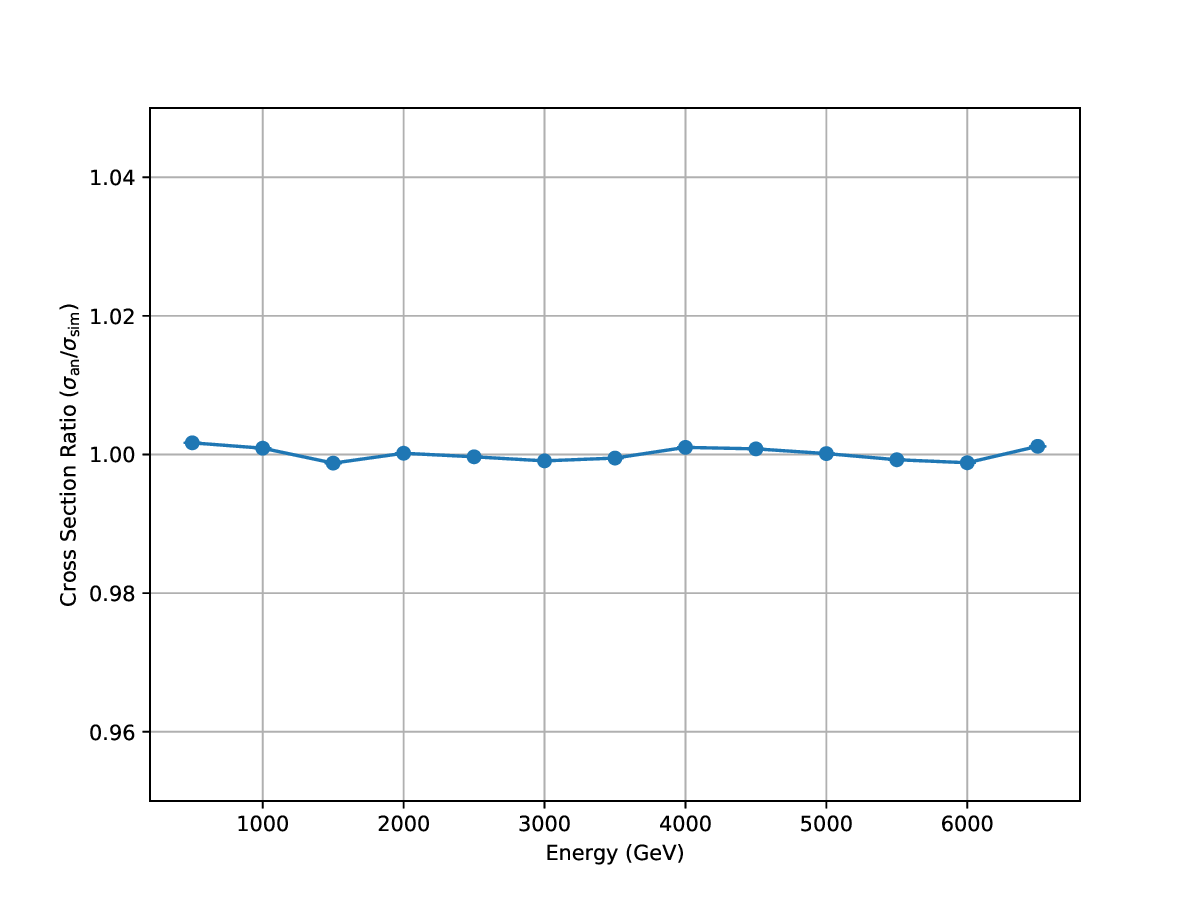}
    \caption{Ratio of $\sigma_{s} / \sigma_{a}$ vs. beam energy $E$ (GeV) for $K_D^+ K_D^- \rightarrow K_D^+ K_D^-$. We can see that the ratio is $1$ within error, thus further validating the four meson term in our model. The points have small error bars that do not appear on the plot, corresponding to the errors in \textsc{MadGraph} simulations.}
    \label{ratio2}
\end{figure}
 The Wess--Zumino--Witten term allows anomalous decay of $\pi^0 \rightarrow \gamma \gamma$. In analogy to the Pion-like DM model, we should retain the same decay rate for the process $\pi_D^0 \rightarrow \gamma_D \gamma_D$ if we use the same parameters. 
\\ The vertex of $\pi^0_D \gamma \gamma$ is
\begin{equation}
   -\frac{\dot\imath c^2 c'^2 g^2_D \kappa^2 \epsilon_{\mu\nu\rho\sigma}(p_1^\sigma - p_2^\sigma)p_3^\rho}{8F\pi^2}
\end{equation}
where $p_1$ and $p_2$ are the 4-momenta of $\gamma$ and $p_3$ is the 4-momentum of the $\pi^0_D$.

 The decay rate for $\pi^0 \rightarrow \gamma \gamma$ is~\cite{scherer2002introduction}
    \begin{equation}\label{eq:4.2.1}
        \Gamma = \frac{\alpha^2 m^3_{\pi 0}}{64\pi^3 F^2}
    \end{equation}
    We should retain the same decay width if we set $F = 0.14$, $m_{\pi_D^0} = 0.135$, and $\alpha = \frac{g_D^2}{4\pi}$:
    $$\Gamma = 3.86459 \times 10^{-9} \ \mathrm{GeV}$$

    The decay width generated by \textsc{MadGraph} is 
    $$\Gamma = 3.865 \times 10^{-9} \pm 5.7 \times 10^{-18} \ \mathrm{GeV}$$
This supports our implementation of the WZW Lagrangian as being correct.
\begin{figure}[htbp]
    \centering
    \includegraphics[scale = 0.8]{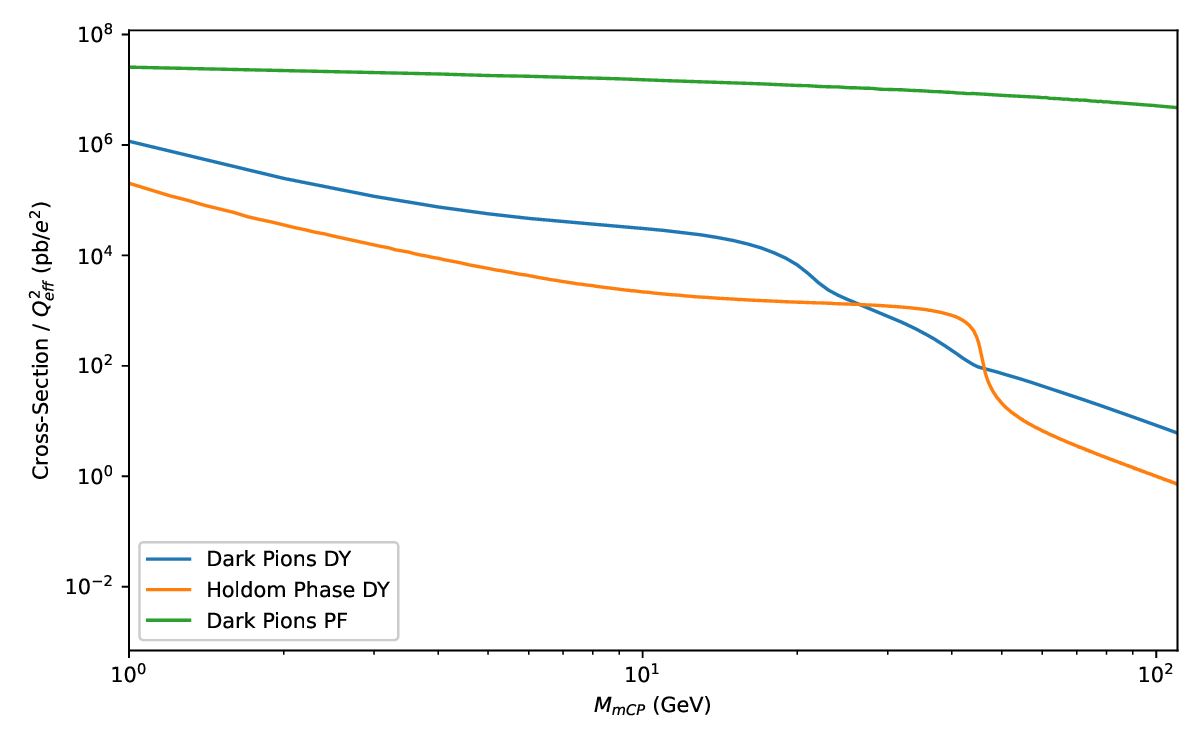}
    \caption{Cross-sections of Dark Pions Drell--Yan, Dark Pions photon fusion, and Holdom phase Drell--Yan for $M_{Z'}$ = $45$~GeV.}
    \label{XSplot}
\end{figure}
For $F=1$~GeV, we plot the cross-section vs. mass of dark pions for the Drell--Yan and photon-fusion cases, this shown in Figure~\ref{XSplot}. We also include the Drell--Yan cross-section for fermionic milli-charged particles coupled to a massless dark photon, also known as the Holdom phase \cite{Holdom:1985ag,Izaguirre_2015} as a benchmark for comparison.  Looking at the Drell--Yan plot, we can see the dark $Z$ resonance in the Drell--Yan cross-section around $m_{\pi_D} = 22.5$~GeV, which is about half the mass of the dark $Z$ boson. However, we also see additional enhancements to the cross-section, which are explained by the usual $Z$-boson resonance. We have picked $M_{Z'} = 45$~GeV as a benchmark; other masses of the dark $Z$ boson will have their resonances at different masses of dark pions. 

As can be seen, at this value of $F$, the photon-fusion plot gives us a higher cross-section. This is due to a difference in its four-momentum dependence compared to Drell--Yan production; the photon-fusion cross-section has a quartic scaling with four-momentum, and Drell--Yan scales quadratically. Due to the high beam energy ($\sqrt{s} = 14$~TeV), we obtain a significantly higher cross-section for the former than for the latter. We also do not see any resonances in the cross-section; rather, it is a smooth curve. This is understandable given that we have not implemented the dark $Z$ or the SM $Z$ term for the WZW term in our model.

For higher values of $F$ we see a rapid drop in the cross-section as it scales with $F$ as $\frac{1}{F^{6}}$. 

\begin{figure}[htbp]
    \centering
    \includegraphics[scale = 0.8]{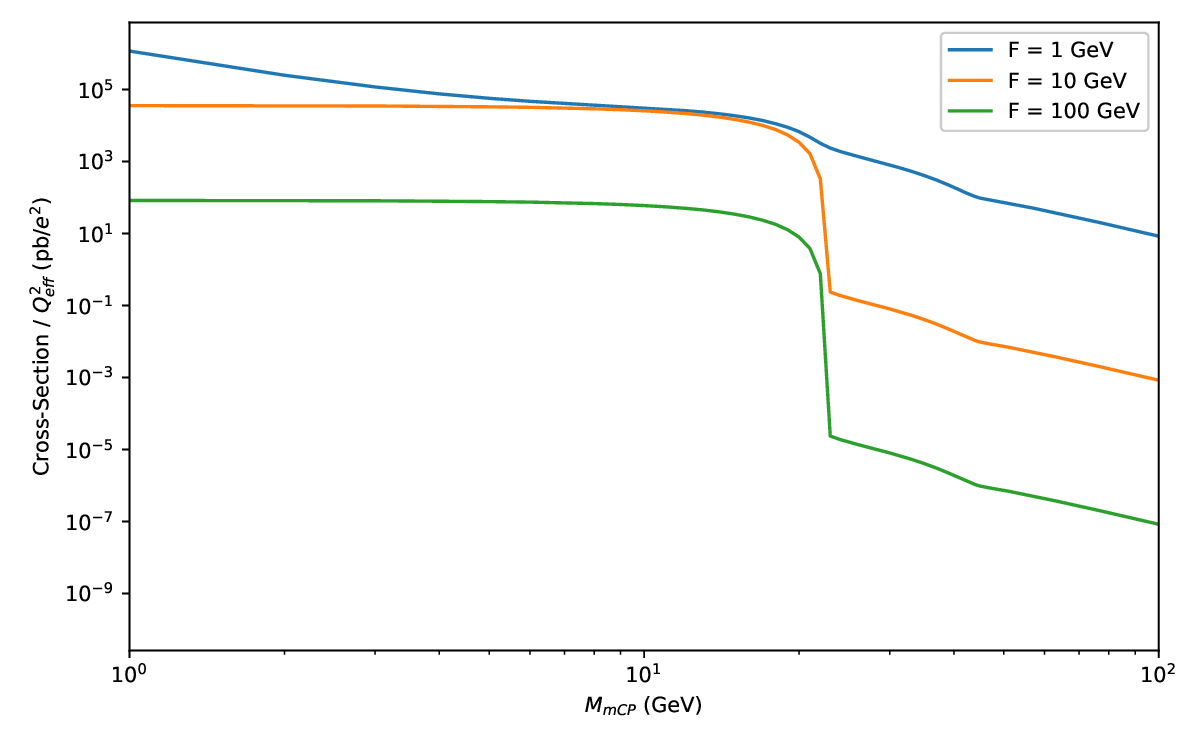}
    \caption{Cross-sections of Drell--Yan production of Dark Pions with different values of $F$ for $M_{Z'}$ = $45$~GeV.}
    \label{diff_f_mz_0.5}
\end{figure}
 As (Eq.~\eqref{eq:2.1.0.5}) only fixes the ratio of $F$ to the dark pion mass, we can consider other values of the dark-pion decay constant $F$. We plot Drell--Yan cross-sections for different values of $F$ in Figure~\ref{diff_f_mz_0.5}.

 Note that for higher values of $F$ the cross-section is much smaller as it scales with $F$ as $\frac{1}{F^{4}}$.
  
We also see a sharp transition in the cross-section when the value of $M_{\mathrm{mCP}}$ is approximately half that of $M_{Z'}$. This sharp transition is driven by a large change in the dark $Z$'s contribution for this process.

\begin{figure}[htbp]
    \centering
    \includegraphics[scale = 0.5]{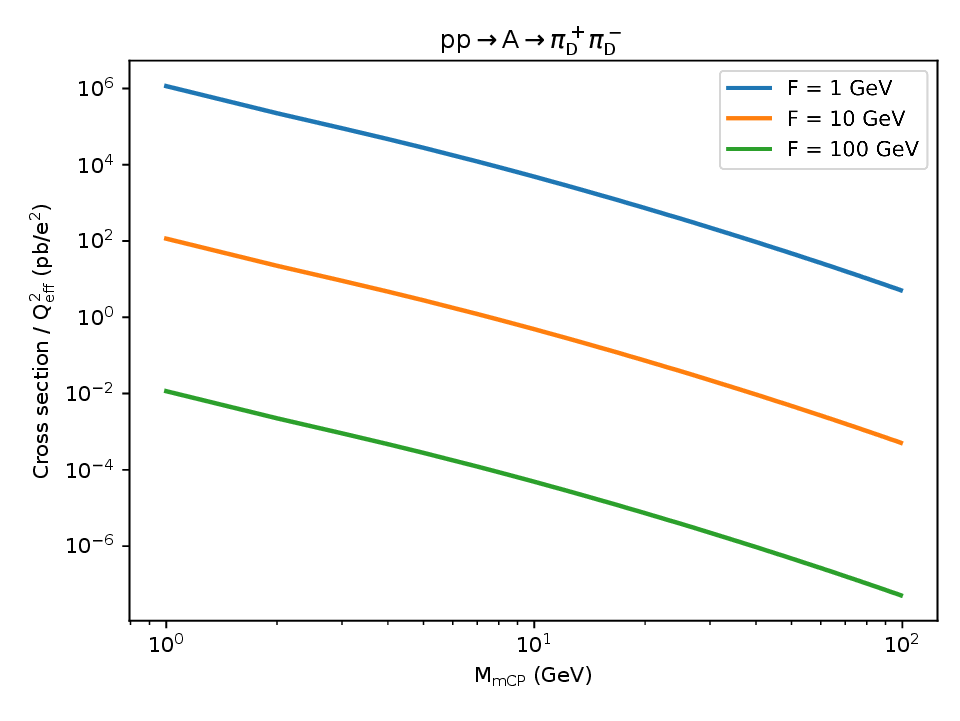} \includegraphics[scale = 0.5]{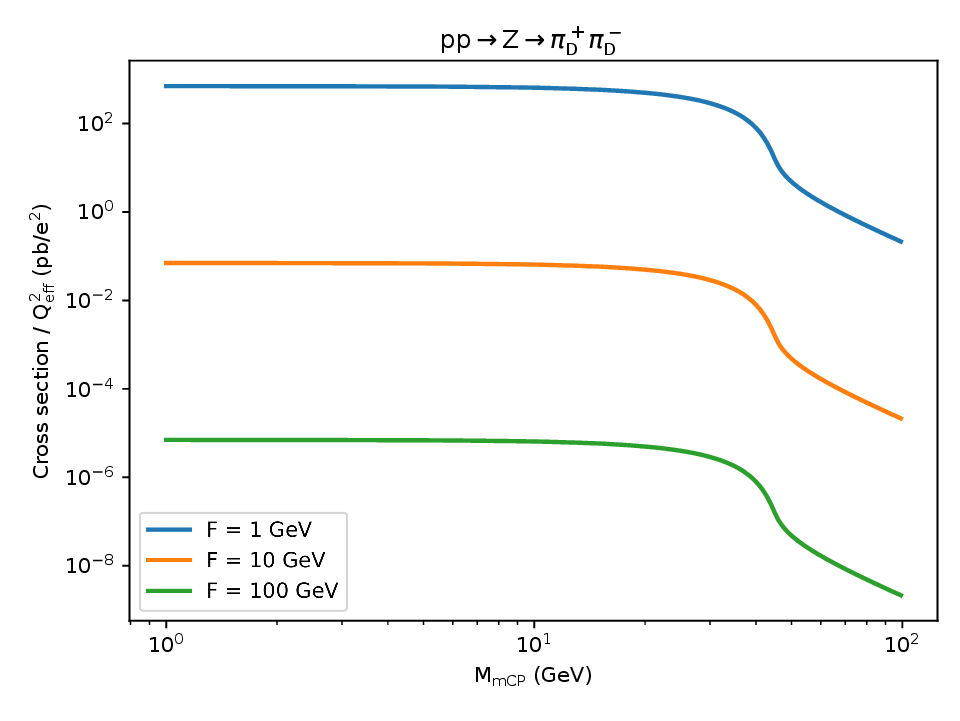} \includegraphics[scale = 0.5]{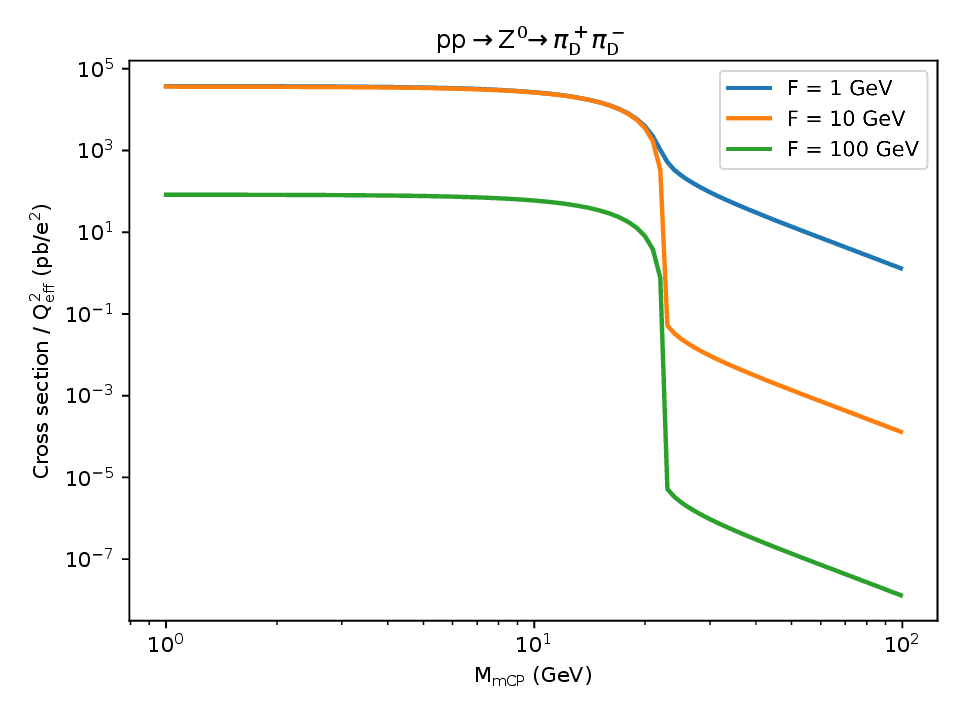}
    \caption{The Drell--Yan production of Dark pions considering only the photon (top), the Z (middle) and the dark Z (bottom).}
    \label{DifContrib}
\end{figure}

To better understand this, we can consider the contributions to the Drell--Yan production of dark pions separately. In Figure~\ref{DifContrib} we plot the Drell--Yan cross section for charged dark pions for each individual gauge field contributing. Modulo the cross terms in the squared amplitude, which do not change the point made, this gives us a better understanding of how each gauge field contributes to the total cross section.

We see that the contribution from the photon decreases more or less monotonically as a function of the dark pion mass. The contribution from the Z is more or less constant until the mass of the dark pion reaches that of the dark Z, then decreases sharply and decreases more or less monotonically afterwards. Finally, the contribution due to the dark Z shows a pattern that is very similar to that of the Z, but with a dramatically sharper drop when the dark pion reaches half the mass of the dark Z.

The behaviour of these cross section is understandable if one considers it carefully. First, a Parton Distribution Function (PDF) is used by MadGraph to model the 4-momentum distribution of the partons in the protons that are collided to generate our dark pions. These PDF will give the distribution of energy and 3-momenta for the quark and anti-quark needed for our process. As can be seen from Figure-4 in Ref. \cite{placakyte2011partondistributionfunctions}, for sea quarks the probability of finding a quark with a given $x$, where $x$ can be understood as the ratio of the parton energy to that of the proton, decreases sharply as the value of $x$ approaches 1. For valence quarks, the story is more complicated as the probability of finding such a quark increases until $x$ reaches a value of about 0.2, corresponding to roughly 1.4 TeV. However, all anti-quarks will be found in the sea and our choice of dark pion mass places us considerably below the peak in $x$ for valence quarks. The probability of finding a quark anti-quark pair with the right energy to form our dark pions will therefore go down monotonically as we increase the mass of the dark pion.    

We can allready see the effect of this lower probability as it is driving down the contribution due to the photon as we increase the dark pion mass. As the value of this mass increases, the more probable low energy quarks drop off, leaving only less probably higher energy quark contributing to the cross section.

Further, for massive gauge fields MadGraph uses a ``dressed" propagator \cite{Alwall_2015_prop} of the form 
\begin{eqnarray*}
    \frac{1}{q^{2}-M_{Z'}^{2}+i\Gamma} 
\end{eqnarray*}
where $\Gamma$ is the width of the gauge field.

As we scan through values of the dark pion mass, we transition from a regime where values of $q^{2}<<M^{2}_{Z}$ or $q^{2}<<M^{2}_{Z'}$ are allowed and the propagator is dominated by $M^{2}_{Z}$ or by $M^{2}_{Z}$ to one where these values of $q^{2}$ are not allowed. In the first of these regions the $Z$ and dark $Z$ contribution are effectively constant as the bulk of the contributions to the integration used to determine the cross section come from quarks with small $q^{2}$ and everything is effectively proportional to $\frac{1}{M^{2}_{Z}}$  or $\frac{1}{M^{2}_{Z'}}$. In the second, the higher probability lower valued $q^{2}$ are no longer kinematically allowed, and the integral only considers $q^{2}$ which are large.  In this region it is dominated by the $q^{2}$ and the cross-section is suppressed. We see both regions and the transition between them in the cross-section shown in Figure~\ref{DifContrib} for both the $Z$ and $Z'$ this happens a value of the dark pion mass that is exactly half their mass. Effectively the dark gauge filed comes on shell.

To understand why the result for the dark Z is considerably sharper then that of the $Z$, we must note that MadGraph dynamically computes the widths of particles involved in all processes. As we transition from a dark pion that is lighter then half the mass of the dark Z to one that is heavier, the dark Z is no longer allowed to decay into a dark pion. This meaningfully changes the width of the dark Z and alters the value used in the propagator during integration. This changes the location of poles and sharply effects the result. To confirm this, we re-ran our calculation using an unphysical fixed width for the dark Z and obtained a contribution that behaved more or less exactly as a ``lighter'' Z would and notably did not have the marked sharp decrease seen.

 The other significant difference between low and high $F$ is that we also see a decline in the cross-section's value in the low $M_{\mathrm{mCP}}$ region of the plot. This change is also driven by the proton's PDF. As mentioned above, the probability of finding a quark and anti-quark with sufficient momentum to produce our milli-charged particle final state is much higher for lower values of $M_{\mathrm{mCP}}$, when coupled with a much higher overall cross-section this drives the observed effect. To confirm that this is the reason for the observed decline, we looked at the same process for $e{+}e^{-}$ in \textsc{MadGraph} and found that the $F=1$~GeV cross-section for that process did not show the same initial decline as seen in Figure~\ref{diff_f_mz_0.5} for protons. 

\begin{figure}[htbp]
    \centering
    \includegraphics[scale = 0.8]{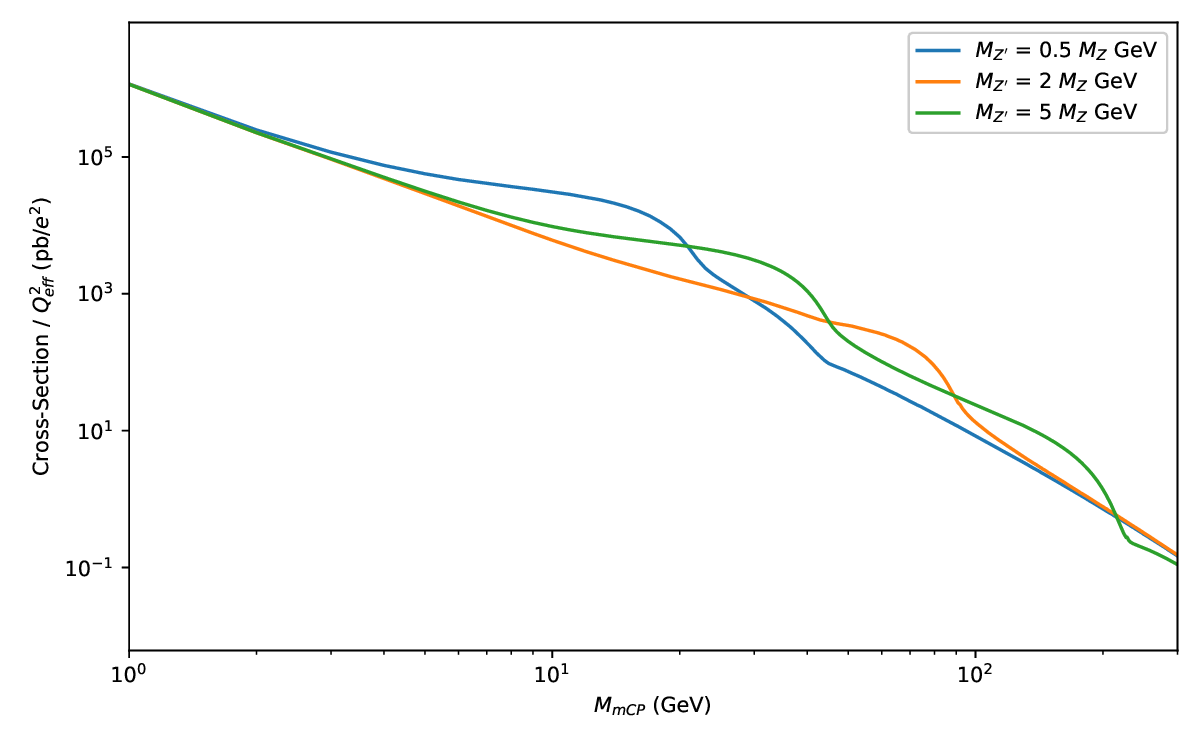}
    \caption{Cross-sections of Dark Pions with different masses of $M_{Z'}$ with $F = 1$~GeV.}
    \label{fig:4.3}
\end{figure}
Figure~\ref{fig:4.3} shows cross-section plots for different masses of $M_{Z'}$. In this plot, we show the cross-sections for $M_{Z'} \sim 22$~GeV, for $M_{Z'} \simeq 45$~GeV, $\sim 91$~GeV for $M_{Z'} \simeq 180$~GeV, and $\sim 228$~GeV for $M_{Z'} \simeq 455$~GeV, these values corresponding to half, double and five times the mass of the SM boson $Z$, respectively. As expected, we see that the dark $Z$ resonance is present at different values of the dark pion masses.
\begin{figure}[htbp]
     \centering
     \includegraphics[scale=0.8]{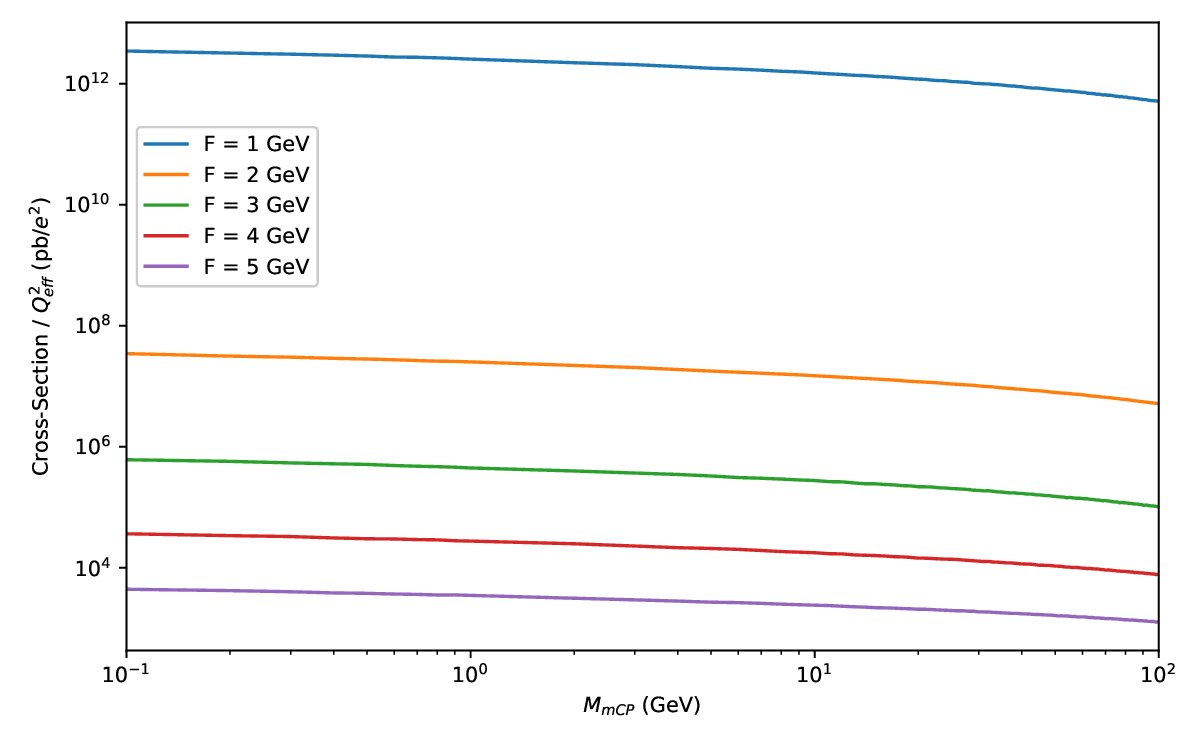}
     \caption{Cross-sections of photon fusion of Dark Pions with different values of $F$,}
     \label{fig:diff_f_pf_xs}
 \end{figure}
 The sharp reduction in the cross-section at higher values of $F$ also strongly suppresses the photon-fusion process. As this process is further kinematically suppressed because of its three-body final state, even if its momentum scaling is quartic, it is relevant only for a narrow band of possible values of F. We show the photon-fusion cross-sections for $F = 1$--$5$~GeV in Figure~\ref{fig:diff_f_pf_xs}.

\subsection{ Sensitivity Contours for DY and PF production}  
\begin{figure}[htbp]
    \centering
    \includegraphics[scale=0.8]{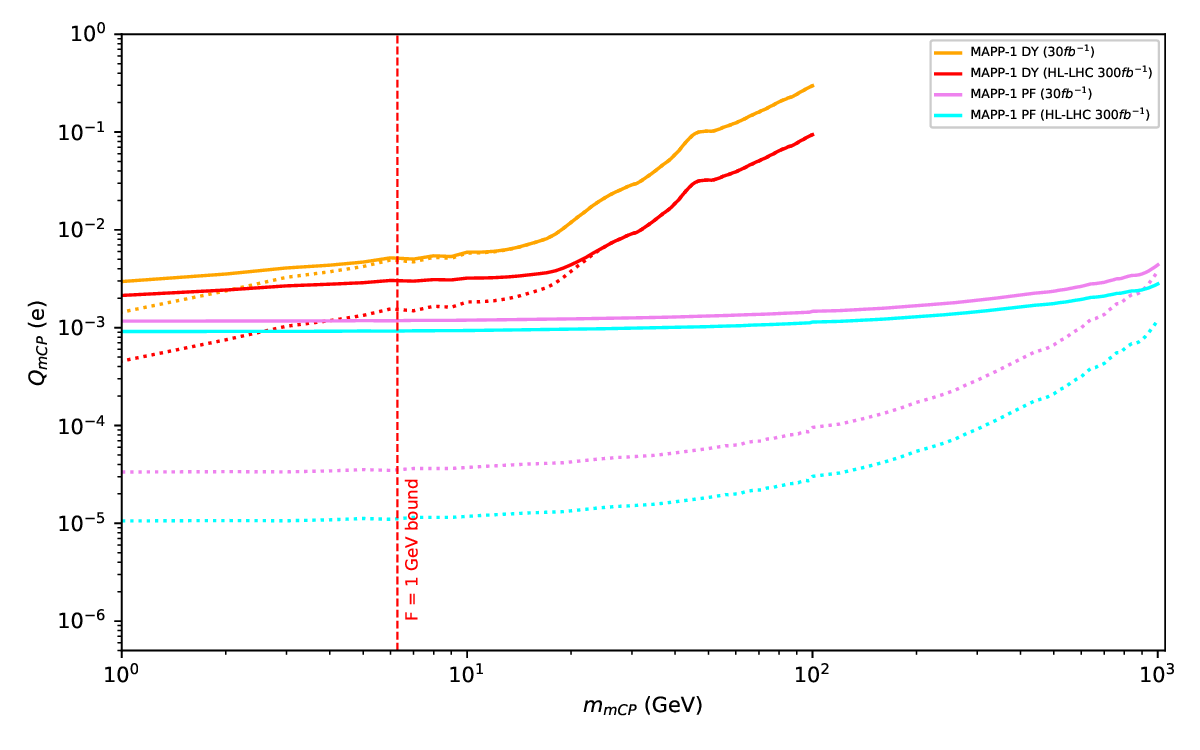}
    \caption{The estimated reach of the MAPP-1 detector for DY pair-produced dark pions and PF production of $3$ dark pions at $\sqrt{s} = 14$~TeV for $30 \textrm{ fb}^{-1}$ and the upcoming HL-LHC, excluded at the $95\%$ CL for $F = 1$ GeV and $M_{Z'} = 0.5 M_Z$. The dotted line represents the sensitivity without any influence of $P$ from Eq.~\eqref{eq:5.26}, while the solid line factors in the detector sensitivity. The vertical line shows the $F=1$~GeV bound as discussed in Equation~\eqref{eq:2.1.0.5}.}
    \label{fig:excl_1}

\end{figure}

To calculate the maximal sensitivity contours, we assume a background of zero and an initial estimate of detector efficiency. We then find the effective charge at which we expect to see three events for the given luminosity as a function of the mass of the dark pion. We then use the following formula to solve for $Q_{\textrm{eff}}$:
    \begin{align*}
        &N_{\sigma} = N_\chi \times A \times P \\ &N_\chi = \sigma \times Q_{\textrm{eff}}^2 \times L_{\textrm{LHCb}}
    \end{align*}
where for IP8, $L_{\textrm{LHCb}} = 30 \textrm{ fb}^{-1}$ and $L_{\textrm{LHCb}} = 300 \textrm{ fb}^{-1}$ for the High-Luminosity phase of the LHC (HL-LHC),
    and  $$A = \frac{\textrm{number of particles that traverse the full MAPP detector}}{\textrm{Total number of particles}}.$$
At the $95\%$ CL, which requires us to have a minimum of $3$ `hits' in the MAPP detector in the background-free scenario, and having $10^6$ events, the expression for $A$ becomes
\begin{equation}\label{eq:5.25}
    A = \frac{3}{N\times 10^6}
\end{equation}
$N$ in the denominator of Eq.~\eqref{eq:5.25} is the number of particles detected. For Drell--Yan, $N=2$, while for photon fusion, $N=3$.
\\ $P$ is the detection probability for a through-going particle, and is defined by~\cite{Kalliokoski:2023cgw}
\begin{equation}\label{eq:5.26}
    P = (1-e^{-N_{\mathrm{PE}}})^n
\end{equation}
where $N_{\mathrm{PE}}$ represents the number of photoelectrons detected, which is proportional to the number of optical scintillation photons reaching the PMT $(N_\gamma)$ and its quantum efficiency (QE). We follow the same method used in Ref.~\cite{Kalliokoski:2023cgw}, considering a QE of $20\%$ for our estimates, which gives us $N_{\mathrm{PE}} \simeq 1.365\times 10^5$ $Q^2$. Equation \eqref{eq:5.26} is a detection probability model based on Poisson photoelectron statistics in a scintillator detector. The formula estimates the probability that an mCP produces a detectable coincidence across n=4 layers, thus helping estimate overall detector efficiency. Its effect can be shown in Figure \ref{fig:excl_1}.

 The procedure for obtaining the acceptances is as follows:
\begin{itemize}
    \item Generate $p p \rightarrow \pi_D^+ \pi_D^-$ and $\gamma \gamma \rightarrow \pi_D^+ \pi_D^- \pi_D^0$ events in \textsc{MadGraph} and \textsc{WHIZARD}~\cite{Kilian_2011}, respectively.
    \item Use an LHE parser to extract the 3-momenta of the particles from the LHE files.
    \item Use the location of the MAPP-1 detector: $97.8$~m away from the IP at an angle of $7.3\degree$, with the full volume of the MAPP-1 detector ($1$~m $\times$ $1$~m $\times$ $3$~m).
    \item Find the number of `hits', i.e, the number of particles with momenta that traverse the full length of the detector for a particular value of the mass of the dark pions.
    \item Use the detection probability in Eq.~\eqref{eq:5.26} to estimate the sensitivity of the detector.
\end{itemize}
We then plot $Q_{\textrm{eff}}$ versus $m_{\mathrm{mCP}}$ to present the bounds of the parameter space we cover. Since we will encounter some statistical fluctuations due to Monte Carlo simulations, we passed our data through a sliding Savitzky--Golay filter~\cite{Savgol} to smooth our plots.

In this study, the projected exclusion limits were estimated under a background (BG)-free scenario. Generally, MAPP-1 is shielded from BGs produced by cosmic rays by an overburden of $110$~m of rock, from low-energy beam-related BGs by $8$~m of concrete, and from collision-related BGs by $\sim 50$~m of rock. The MAPP-1 detector also has its hermetic veto system and quadruple-coincidence requirements for the signal, giving it an advantage of utilizing several BG rejection techniques.

\begin{figure}[htbp]
  \centering

  \begin{subfigure}{0.7\textwidth}
    \centering
    \includegraphics[width=\linewidth]{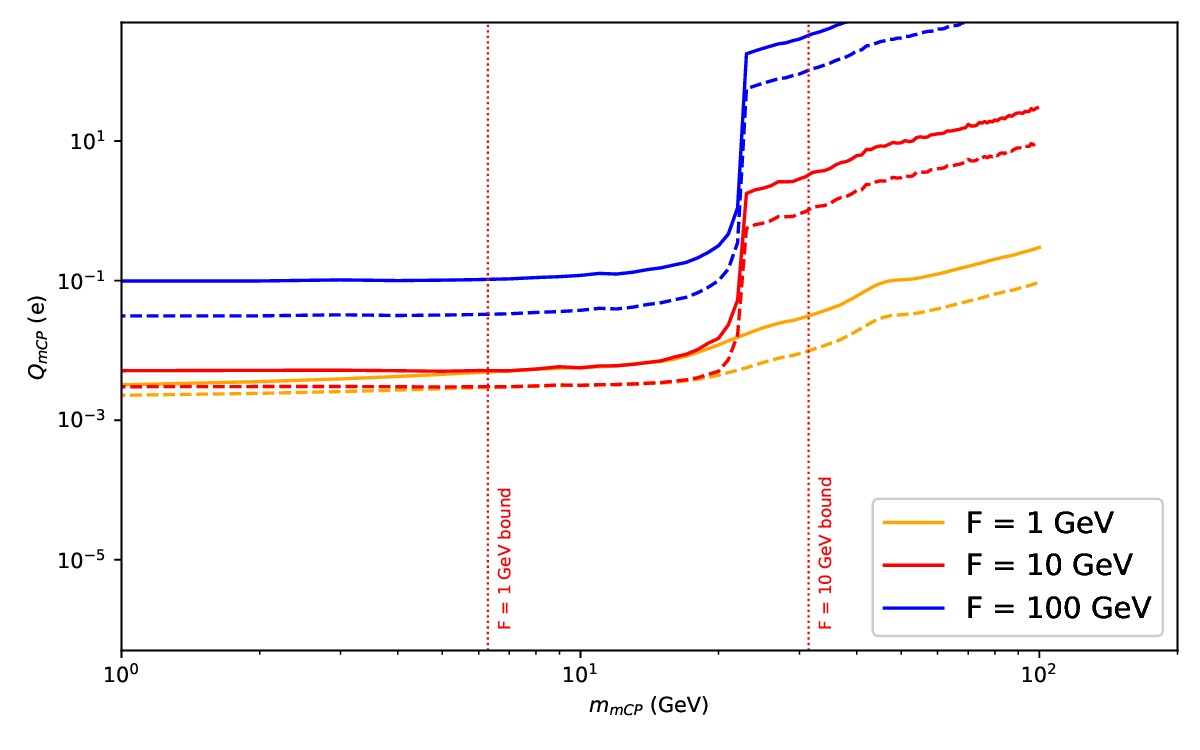}
    \caption{}
    \label{fig:excl_full_mz_1}
  \end{subfigure}

  \begin{subfigure}{0.7\textwidth}
    \centering
    \includegraphics[width=\linewidth]{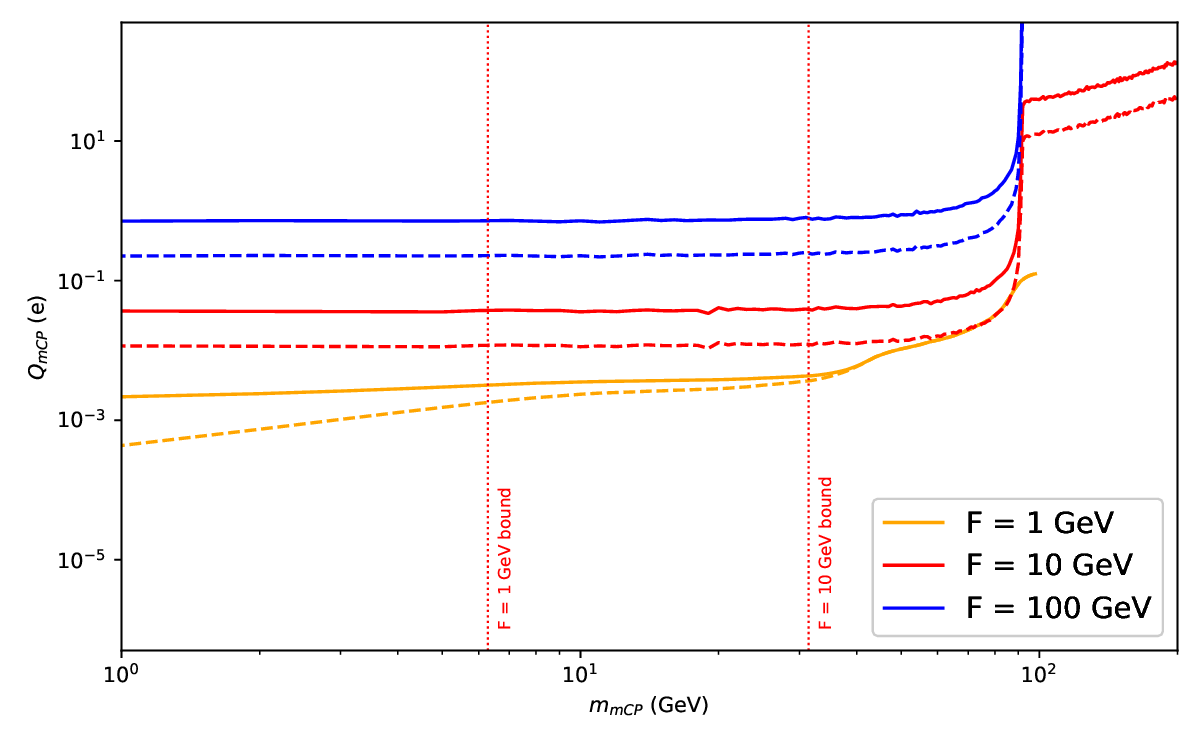}
    \caption{}
    \label{fig:excl_full_mz_2}
  \end{subfigure}

  \caption{The estimated reach of the MAPP-1 detector for DY pair-produced dark pions at $\sqrt{s} = 14$~TeV for $30 \textrm{ fb}^{-1}$ and the upcoming HL-LHC ($300 \textrm{ fb}^{-1}$), excluded at the $95\%$ CL for different values of $M_{Z'}$. (a) $M_{Z'} = 0.5 M_Z \approx 45$~GeV; (b) $M_{Z'} = 2 M_Z \approx 180$~GeV. The yellow, blue, and green lines are for $F = 1$, $F = 10$, and $F = 100$~GeV, respectively. The solid lines are for $30 \textrm{ fb}^{-1}$, while the dashed lines are for the HL-LHC ($300 \textrm{ fb}^{-1}$). The red dotted vertical lines are the model-dependent cosmological limits for $F = 1$~GeV and $F = 10$~GeV. In these limits we have used Equation \eqref{eq:5.26} to model detector response.
  }
  \label{fig:fig}
\end{figure}

Figure~\ref{fig:excl_1} shows the bounds for the production of dark pions for both Drell--Yan and photon-fusion processes for $F = 1$~GeV and $M_{Z'} = 45$~GeV. Sensitivity plots of different values of $F$ and $M_{Z'}$ are shown in Figure~\ref{fig:excl_full_mz_1},\ref{fig:excl_full_mz_2}. We can see a sharp decrease in sensitivity at $\sim 0.5 M_{Z'}$ due to the resonance in the cross-section shown in Figure~\ref{diff_f_mz_0.5}. This gives us a reasonable parameter space until the resonance $(\sim 22.5$~GeV for Figure~\ref{fig:excl_full_mz_1} and $\sim 90$~GeV for Figure~\ref{fig:excl_full_mz_2}$)$. It is also useful to note that the cosmological limit for $F = 10$~GeV is within a desirable sensitivity when $M_{Z'} = 2 M_Z$, and so we can obtain sensible limits with higher $M_{Z'}$ for higher values of $F$.

\begin{figure}[htbp]
    \centering
    \includegraphics[scale=0.8]{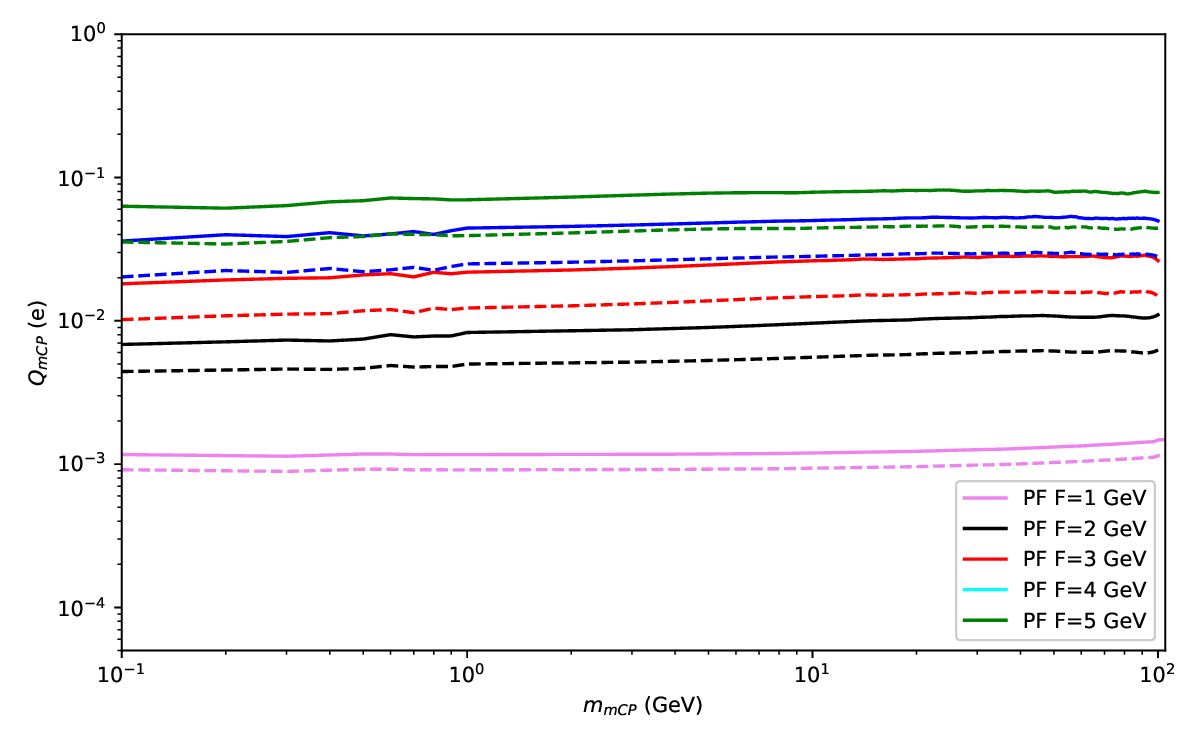}
    \caption{The estimated reach of the MAPP-1 detector for photon-fusion production of dark pions at $\sqrt{s} = 14$~TeV for $30\textrm{ fb}^{-1}$ and the upcoming HL-LHC ($300 \textrm{ fb}^{-1}$), excluded at the $95\%$ CL for different values of $F$. The solid lines are for $30 \textrm{ fb}^{-1}$, while the dashed lines are for the HL-LHC ($300 \textrm{ fb}^{-1}$). The pink, black, red, blue, and green lines are for $F = 1, \ 2,\ 3,\ 4,\ 5 $~GeV, respectively.}
    \label{fig:PF_excl}
\end{figure}

\FloatBarrier
The photon-fusion sensitivity limits are shown in Figure~\ref{fig:PF_excl}, we are only covering values of $F = 1$--$5$~GeV, as the photon-fusion cross-section scales with $\sigma \sim 1/F^6$, and so the cross-section diminishes with higher values of $F$.

\begin{figure}[htbp]
  \centering

  \begin{subfigure}[b]{0.6\linewidth}
    \centering
    \includegraphics[width=\linewidth]{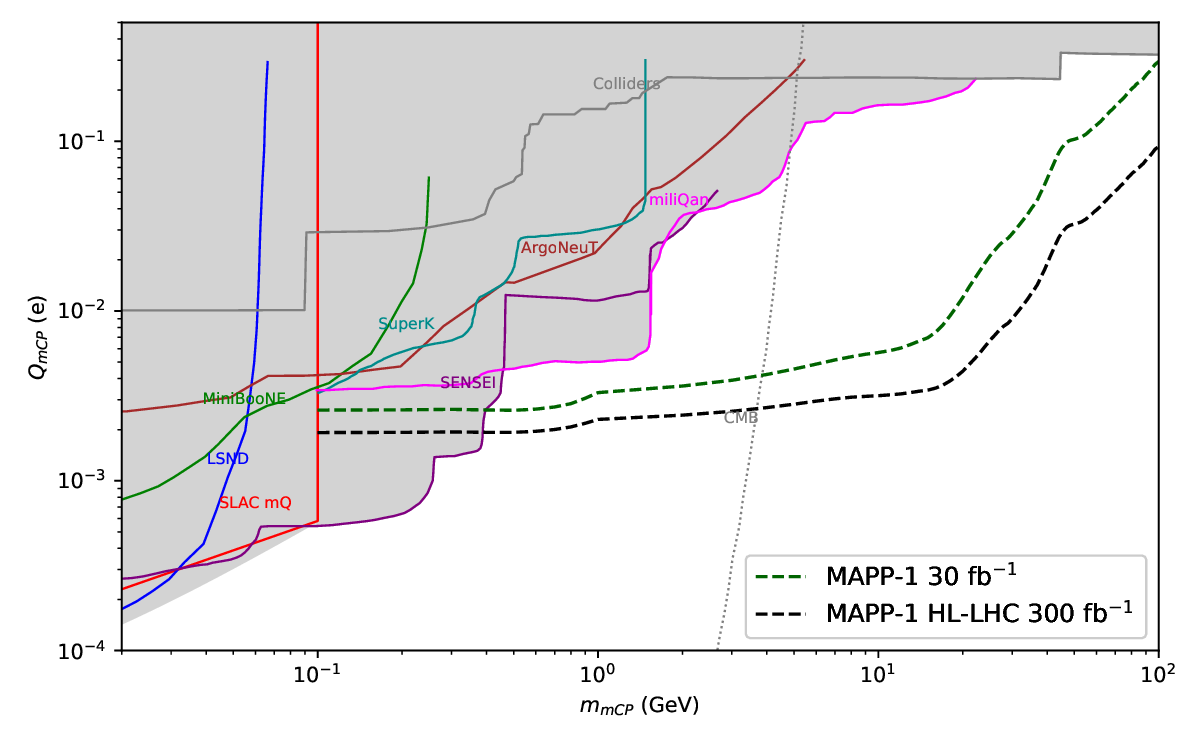}
    \caption{}
    \label{fig:excl_full1}
  \end{subfigure}

  \begin{subfigure}[b]{0.6\linewidth}
    \centering
    \includegraphics[width=\linewidth]{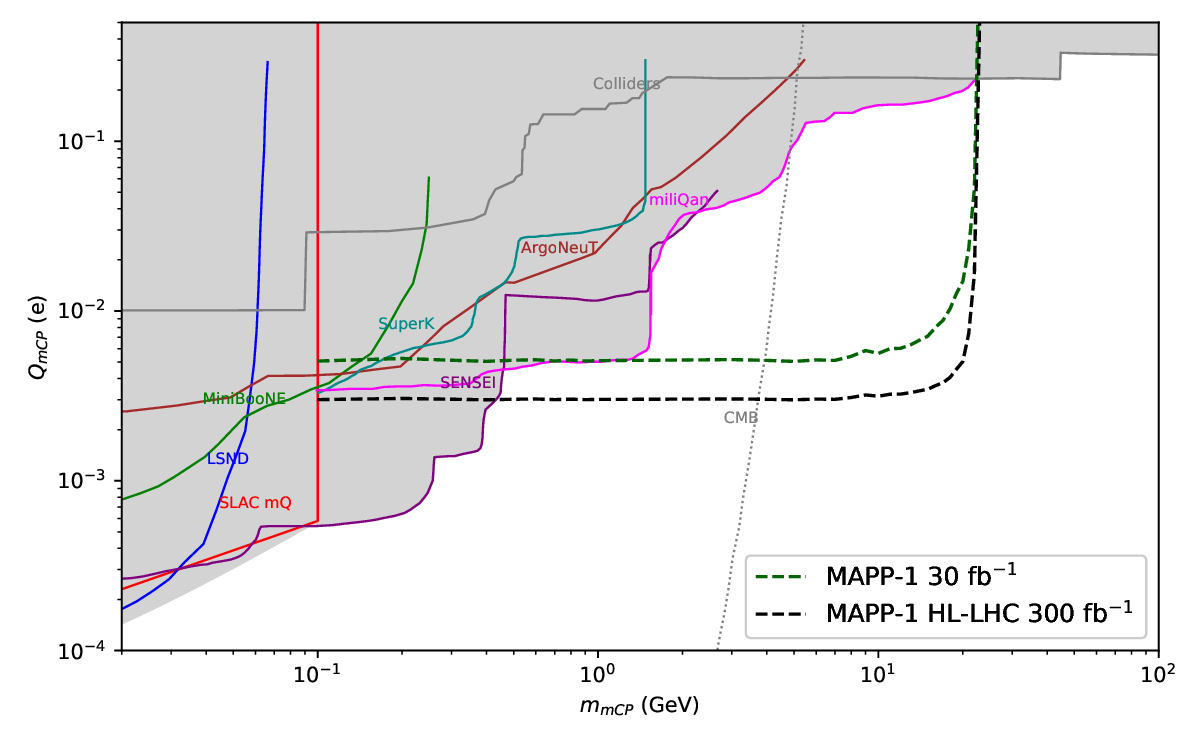}
    \caption{}
    \label{fig:excl_full10}
  \end{subfigure}

  \begin{subfigure}[b]{0.6\linewidth}
    \centering
    \includegraphics[width=\linewidth]{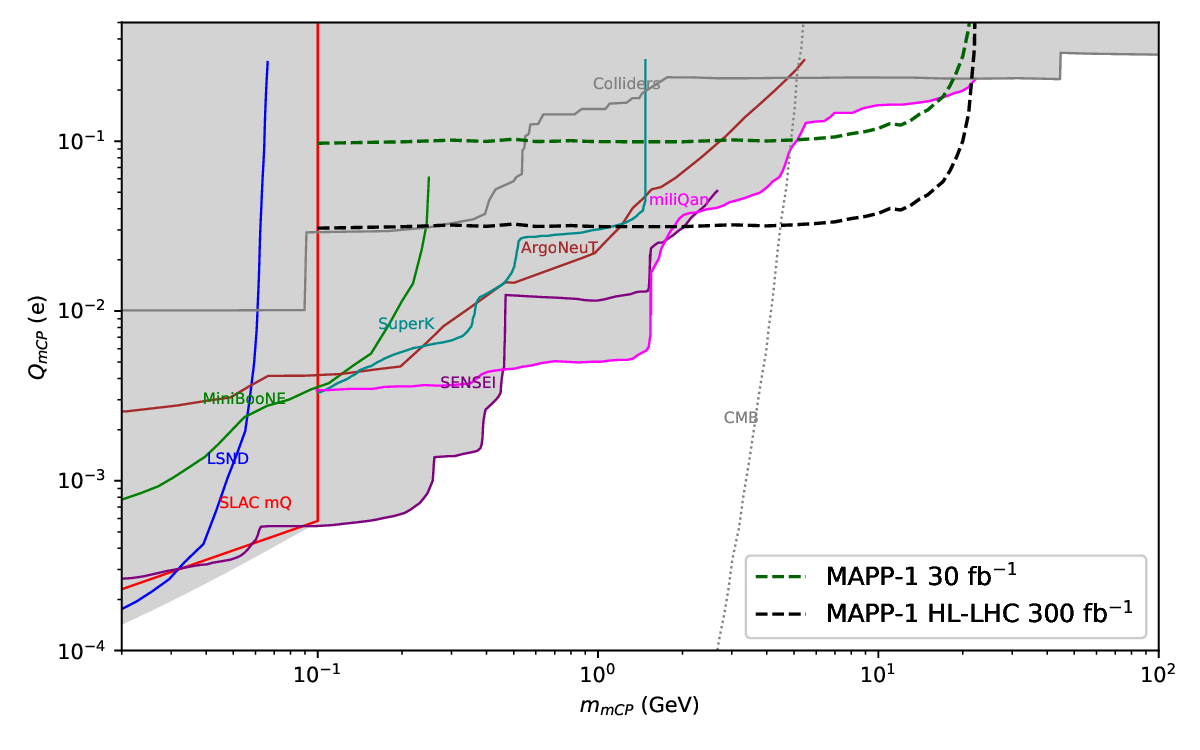}
    \caption{}
    \label{fig:excl_full100}
  \end{subfigure}

  \caption{Estimated reach of the MAPP-1 detector for DY pair-produced dark pions 
  at $\sqrt{s} = 14$~TeV for 30~$\textrm{fb}^{-1}$ and the HL-LHC ($300 \textrm{ fb}^{-1}$), excluded at the $95\%$ CL with $M_{Z'} = 0.5 
  \ M_Z$. (a) $F = 1$ GeV; (b) $F = 10$ GeV; (c) $F = 100$ GeV.
  Also shown are $100\%$ efficiency fermionic mCP bounds from other experiments for context~\cite{SLAC,Davidson_2000,Magill_2019,argoNeuT1,milliQan,Plestid_2020,Bays_2012,Kovetz_2018}.} 
  \label{fig:excl_full}
\end{figure}

Since there are no searches for milli-charged dark pseudoscalars at collider level in the literature, we have included the sensitivity plots of searches for milli-charged fermions (Holdom phase) for various experiments in Figure~\ref{fig:excl_1} to show our bounds in context. It is useful to note that the bounds for other experiments are done as a full study, while ours are BG-free. We are also stressing that our bounds cannot be compared with the milli-charged fermion studies, and they are shown here just for context.

\FloatBarrier
\section{Conclusions}\label{Section:Conclusions}

We have explored the prospects for detecting pion-like dark pseudoscalar dark matter with the MoEDAL-MAPP experiment. As shown in Figure~\ref{fig:excl_full}, our bounds open up previously unconstrained regions of parameter space for dark scalars. We examined two production channels --- Drell--Yan and photon fusion --- the latter driven by the distinctive Wess--Zumino--Witten term, a hallmark of this model. This term not only sets the model apart but also enables pion-like dark matter to be a viable candidate, as argued in Ref.~\cite{Hochberg_2015}.

The unique kinematics of the three-body final state from photon fusion could allow it to be distinguished experimentally from other modes such as Drell--Yan. While no other milli-charged-particle searches at accelerators have yet probed pion-like dark matter, our results can serve as a benchmark for future studies. In the low-mass regime, mesonic decays from Drell--Yan production dominate --- a feature we plan to investigate further, given the potentially intriguing phenomenology.

Because cross-sections in this model depend on momentum, high-momentum mCPs may deposit energy differently in the detector, offering a possible boost in sensitivity. Finally, since our present sensitivity estimates assume a background-free detector, the natural next step is a detailed \textsc{GEANT4}~\cite{AGOSTINELLI2003250} simulation of the full MAPP-1 setup to refine --- and perhaps extend --- our reach.

\acknowledgments
We acknowledge support by NSERC via Grant Number SAPPJ-2023-00043. P.P.O. and S.A. acknowledge support from the Faculty of Science at the University of Regina (startup). P.P.O. thanks Gojko Vujanovic for useful discussions. M.S. acknowledges support by the Generalitat Valenciana via the APOSTD Grant No. CIAPOS/2021/88.


\newpage

\bibliographystyle{JHEP}
\bibliography{bibliography/ref}   

\end{document}